\documentclass{article}

\usepackage{amsmath}
\usepackage{amsfonts}
\usepackage{natbib}
\usepackage[colorlinks,citecolor=blue,urlcolor=blue,filecolor=blue,backref=page]{hyperref}
\usepackage{graphicx}
\usepackage{mathtools}
\usepackage{algorithm}
\usepackage{algpseudocode}
\usepackage{booktabs}
\usepackage{authblk}
\usepackage{geometry}

\newcommand{\E}{\mbox{$\textrm{\textup{E}}$}}
\newcommand{\var}{\mbox{$\textrm{\textup{var}}$}}

\newcommand{\ds}{\displaystyle}

\newcommand{\dg}{\mbox{$\textrm{\textup{dg}}$}}
\newcommand{\diag}{\mbox{$\textrm{\textup{diag}}$}}
\newcommand\norm[1]{\left\lVert#1\right\rVert}
\newcommand{\appropto}{\mathrel{\vcenter{\offinterlineskip\halign{\hfil$##$\cr\propto\cr\noalign{\kern2pt}\sim\cr\noalign{\kern-2pt}}}}}
\newcommand{\CC}{C\nolinebreak\hspace{-.05em}\raisebox{.4ex}{\tiny\bf+}\nolinebreak\hspace{-.10em}\raisebox{.4ex}{\tiny\bf +}}

\def\vectorfontone{\bf}
\def\vzero{{\vectorfontone 0}}

\def\vb{{\vectorfontone b}
\def\vc{{\vectorfontone c}}}
\def\vd{{\vectorfontone d}}

\def\vm{{\vectorfontone m}}

\def\vs{{\vectorfontone s}}
\def\vt{{\vectorfontone t}}

\def\vv{{\vectorfontone v}}

\def\vectorfonttwo{\boldsymbol}

\def\vdelta{{\vectorfonttwo \delta}}
\def\vDelta{{\vectorfonttwo \Delta}}

\def\vmu{{\vectorfonttwo \mu}}

\def\mSigma{{\vectorfonttwo \Sigma}}
\def\vtheta{{\vectorfonttwo \theta}}
\def\vTheta{{\vectorfonttwo \Theta}}

\def\matrixfontone{\bf}
\def\mA{{\matrixfontone A}}

\def\mC{{\matrixfontone C}}

\def\mJ{{\matrixfontone J}}

\def\matrixfonttwo{\boldsymbol}

\def\mSigma{{\matrixfonttwo \Sigma}}

\def\sD{{\mathcal D}}
\def\sN{{\mathcal N}}

\title{Skew-Normal Posterior Approximations}

\author[1]{Jackson Zhou}
\author[1, 2]{Clara Grazian}
\author[1, 3]{John T. Ormerod}

\affil[1]{\footnotesize School of Mathematics and Statistics, The University of Sydney, NSW 2006, Australia}
\affil[2]{\footnotesize ARC Training Centre in Data Analytics for Resources and Environments (DARE), The University of Sydney, NSW 2006, Australia}
\affil[3]{\footnotesize The work of JO was supported by an Australian Research Council Discovery Project Grant (DP210100521)}

\begin{document}

\maketitle

\begin{abstract}
Many approximate Bayesian inference methods assume a particular parametric form for approximating the posterior distribution.
A multivariate Gaussian distribution provides a convenient density for such approaches; examples include the Laplace, penalized quasi-likelihood, Gaussian variational, and expectation propagation methods.
Unfortunately, these all ignore the potential skewness of the posterior distribution.
We propose a modification that accounts for skewness, where key statistics of the posterior distribution are matched instead to a multivariate skew-normal distribution.
A combination of simulation studies and benchmarking were conducted to compare the performance of this skew-normal matching method (both as a standalone approximation and as a post-hoc skewness adjustment) with existing Gaussian and skewed approximations.
We show empirically that for small and moderate dimensional cases, skew-normal matching can be much more accurate than these other approaches.
For post-hoc skewness adjustments, this comes at very little cost in additional computational time.
\end{abstract}

\section{Introduction}\label{sec:introduction}

Fast, scalable approximations to posterior distributions have been a staple of Bayesian inference when dealing with big data \citep{graves2011practical,zhou2020variational}, especially when Markov Chain Monte Carlo (MCMC) is too costly.
Approximate Bayesian inference (ABI) is a class of alternative methods that includes the Laplace approximation, 
the fully exponential (or improved) Laplace approximation \citep{tierney_kadane_1986}, variational Bayes and related methods, 
and expectation propagation \citep{minka_2001}.
These alternative methods are often deterministic, based on approximating the posterior distribution via a standard distribution, 
and often have a significant advantage in speed over standard MCMC techniques when working with large datasets.

ABI methods typically perform optimization, usually by minimizing the distance between the approximation and the posterior distribution, in place of using sampling as in standard MCMC methods, and, in this way, they cut down on the majority of computational time.
This major benefit is reflected by the extensive application of ABI methods for inference in a wide range of fields \citep{Blei2017}.
However, this increase in speed is clearly also compromised by a decrease in accuracy, as ABI-based approximations can be only as accurate as the distribution used to approximate the posterior.
If the standard distribution used is not a good fit for the posterior, then the approximation can be poor.
In particular, ABI methods mostly base their posterior approximations on the parametric form of a multivariate Gaussian density, in order to take advantage of the asymptotic normality of the posterior distribution guaranteed by the Bernstein-von Mises theorem \citep{Gelman1995}.
While the approximations stemming from these Gaussian-based methods are fast and reasonable, the quality of these approximations will inevitably be questionable when the posterior is not similar in shape to the multivariate Gaussian density.
For example, \cite{tierney_kadane_1986} note that the principal regularity condition required for reasonable Laplace approximations is that the posterior is dominated by a single mode.
Even if the posterior is unimodal, skewness in the posterior density can lead to sub-par fits, as the Gaussian density is not skewed.
This can potentially become a problem when the sample size is small compared to the number of parameters in the model.
For instance, \cite{fong_rue_wakefield_2009} indicate that integrated nested Laplace approximations for generalized linear mixed models tend to be less accurate for binomial data with small sample sizes, which result in skewed posterior distributions.

A recent advancement by \cite{durante2023skewed} addresses this problem by way of a skewed Bernstein-von Mises theorem.
This asymptotic approximation takes the form of a
generalized skew-normal (GSN) distribution \citep{genton2005generalized}.
This results in plug-in-based (using the maximum a posteriori estimate) skew-model (SM) approximations which converge to the posterior distribution faster than those of the classical Bernstein-von Mises theorem, by an order of $1/\sqrt{n}$.
While this method is a very important theoretical contribution and is promising from an accuracy perspective, it is not without its practical drawbacks.
The GSN distribution does not have closed forms expressions for the marginal distributions, nor any of its moments necessitating sampling schemes to 
approximate these quantities (when needed) adding to its computational cost.

Under these considerations, this paper aims to improve on conventional Gaussian-based methods for approximating posterior distributions, by taking skewness into account in such a way that posterior inference is kept simple and practical.
Although many skew-normal distributions now exist \citep{genton_2004}, we propose that the posterior density is approximated by the original multivariate skew-normal (MSN) distribution described in-depth by \cite{azzalini_capitanio_2014}, to take advantage of the extensive theory that has already been developed.
In this new method which we will call skew-normal matching, key statistics of the posterior are initially estimated using existing methods, and the MSN approximation is then constructed by matching these statistics with those of the MSN density.
A number of variants are explored, with each approach offering an intuitive way to take skewness into account when the posterior is not well-behaved.
The performance of each variant is then compared with existing Gaussian approaches and the SM approximation of \cite{durante2023skewed} across a range of simulation and benchmark settings that focus on small and moderate dimensional settings.

The outline of the paper is as follows.
In Section \ref{sec:background}, the multivariate skew-normal distribution is introduced and its existing applications in approximating probability distributions are discussed.
In Section \ref{sec:matching}, a detailed description is given of the skew-normal matching method; four separate matching schemes are covered, with each scheme based on a combination of posterior statistics.
In Section \ref{sec:settings}, the simulation and benchmark settings that the following two sections use are described.
In Section \ref{sec:standalone}, the performance of skew-normal matching as a standalone approximation is compared to that of existing methods, across both simulated data and selected benchmark datasets.
In Section \ref{sec:post-hoc}, post-hoc alternatives of the skew-normal matching method are considered, with performance being evaluated across both simulated data and selected benchmark datasets.
In Section \ref{sec:summary}, the results and future directions are discussed.
Finally, the supplementary material contains derivations and additional plots.

This paragraph summarizes the definitions, conventions, and notation used in this paper.
Let $\sD$ denote the set of observed variables, i.e., the data.
Let $\mathcal{S}_+^{p}$ denote the set of all $p\times p$ positive definite matrices.
Let $\phi_p(\vtheta;\vmu,\mSigma)$ denote the density of the $\sN_p(\vmu,\mSigma)$ distribution evaluated at $\vtheta$ with mean $\vmu\in{\mathbb R}^p$ and covariance $\mSigma\in\mathcal{S}_+^{p}$, and $\Phi$ denote the cumulative distribution function of the standard Gaussian distribution.
The $\odot$ symbol in the exponent denotes the Hadamard power, the operation where all elements in a matrix are raised to the same power.
If $\mA$ is a square matrix then $\dg(\mA)$ is the vector consisting of the diagonal
elements of $\mA$.
If $\vb$ is a vector then $\diag(\vb)$ is a diagonal matrix with diagonal elements $\vb$.
Functions applied to vectors are interpreted as being applied element-wise.

\section{The multivariate skew-normal distribution}\label{sec:background}

While many results based on the MSN distribution are presented using the original parameterization by \cite{azzalini_1996}, we choose a common alternate parameterization that simplifies the exposition.
For a $p$-dimensional random vector $\vTheta$ we will work with the skew-normal density of the form
\begin{align*}
p(\vtheta)=2\cdot\phi_p(\vtheta;\vmu,\mSigma)\cdot\Phi(\vd^\top(\vtheta - \vmu)),
\end{align*}
where $\vmu\in\mathbb{R}^p$, $\mSigma\in\mathcal{S}_+^{p}$, and $\vd\in\mathbb{R}^p$ are the location, scale, and skewness parameters respectively.
In this case, we write $\vTheta\sim\text{SN}_p(\vmu,\mSigma,\vd)$.
An account of the properties of this distribution are given in detail by \cite{azzalini_capitanio_2014}.
Of importance in this paper are the expectation and variance, which are 
\begin{align*}
    \E(\vTheta) = \vmu+\sqrt{\frac{2}{\pi}}\vdelta
    \quad\text{and}\quad
    \var(\vTheta)=\mSigma - \frac{2}{\pi} \vdelta\vdelta^\top
\end{align*}
respectively where $\vdelta = \mSigma\vd/\sqrt{1+\vd^\top\mSigma\vd}$.
Given $\vdelta$ and $\mSigma$, the vector $\vd$ can be recovered via the identity
$\vd = \mSigma^{-1}\vdelta/\sqrt{1-\vdelta^\top\mSigma^{-1}\vdelta}$ leading
to the constraint
\begin{align}\label{eq:constraint}
\vdelta^\top\mSigma^{-1}\vdelta < 1.
\end{align}
The cumulant generating function can be shown to be
$K(\vt)=\log2+\vmu^\top\vt+\vt^\top\mSigma\vt/2 +\log\Phi(\vdelta^\top\vt)$,
from which it can be shown that the vector of third-order unmixed central moments (TUM) is given by
\begin{align}\label{eq:msntum}
    \text{TUM}(\vTheta)=\frac{\sqrt{2}(4-\pi)}{\pi^{3/2}}\vdelta^{\odot3}.
\end{align}
Together, (\ref{eq:constraint}) and (\ref{eq:msntum}) imply a constraint on the allowable `size' of the TUM vector admittable by the MSN distribution.

As a natural extension to the multivariate Gaussian distribution, the MSN distribution lends itself particularly well to approximating probability distributions which are potentially skewed; \cite{azzalini_capitanio_2014} give a brief outline of existing frequentist work using skew-normal approximations in their applications chapter.
In particular, skew-normal approximations to the discrete binomial, negative binomial, and hypergeometric distributions via moment matching have been investigated by \cite{chang2008note}.
Furthermore, unobserved continuous covariates were modelled using the skew-normal distribution by \cite{guolo2008flexible} in the analysis of case-control data.
Additionally, \cite{gupta2003density} have used the skew-normal functional form as the main component of their Edgeworth-type density expansions, in order to account for skewness in the density.

\section{The skew-normal matching method}\label{sec:matching}

In the skew-normal matching method, key statistics of the posterior distribution (for which the density is known) are estimated and then matched with the MSN density to construct a suitable skew-normal posterior approximation.
When the parameter of interest $\vtheta$ has $p$ components, the MSN approximation has a total of $2p+p(p+1)/2$ parameters to vary.
Four matching schemes were devised, each of which matches $2p+p(p+1)/2$ posterior statistics with those of the MSN density, so that there is at most one solution in general.
These are summarized in the following subsections.

\subsection{Moment matching}

In the moment matching (MM) scheme, the mean $\widetilde{\vmu}\in\mathbb{R}^p$ and covariance matrix $\mC\in\mathcal{S}_+^{p}$, along with the TUM vector $\vs\in\mathbb{R}^p$ of the posterior distribution $p(\vtheta|\sD)$ are estimated.
These are then matched to the MSN density to form an approximation of the posterior.
This is similar in nature to the work of \cite{chang2008note}, where common distributions were moment matched to a skew-normal distribution in the univariate case.
The practicality of this approach is limited as posterior moments are often not readily available.
The final system of matching equations for the moment matching scheme are given by
\begin{subequations}\label{eq:mmatch}
\begin{align}
    \widetilde{\vmu}
    & = \vmu+\sqrt{\frac{2}{\pi}}\vdelta, \label{eq:mma} 
    \\
    \mC 
    & = \mSigma-\frac{2}{\pi}\vdelta\vdelta^\top, \label{eq:mmb} 
    \\
    \vs 
    & = \frac{\sqrt{2}(4-\pi)}{\pi^{3/2}}\vdelta^{\odot3},\ \mbox{and} \label{eq:mmc} \\
    \vdelta
    & = \frac{\mSigma\vd}{(1+\vd^\top\mSigma\vd)^{1/2}}. \label{eq:mmd}
\end{align}
\end{subequations}
The derivations are provided in the supplementary material.
It is straightforward to obtain the values $\vdelta$, $\vmu$, and $\mSigma$ of the approximating MSN distribution.
Once $\vdelta$ and $\mSigma$ are known, $\vd$ can also be recovered easily.
The steps of the moment matching scheme are outlined in Algorithm \ref{alg:mm}.

\begin{algorithm}
\caption{Skew-normal matching method: moment matching}\label{alg:mm}
\begin{algorithmic}[1]
\Require $\widetilde{\vmu}\in\mathbb{R}^p,\ \mC\in\mathcal{S}_+^{p},\ \vs\in\mathbb{R}^p$
\State $\ds \vdelta^* \gets \left[\frac{\pi^{3/2}}{\sqrt{2}(4-\pi)}\right]^{1/3}\cdot \vs^{\odot1/3}$
\State $\ds \vmu^* \gets \widetilde{\vmu}-\sqrt{\frac{2}{\pi}}\vdelta^*$
\State $\ds \mSigma^* \gets \mC+\frac{2}{\pi}\vdelta^*(\vdelta^*)^\top$
\State $\ds \vd^* \gets (\mSigma^*)^{-1}\vdelta^*/\sqrt{1-(\vdelta^*)^\top(\mSigma^*)^{-1}\vdelta^*}$ \\
\Return $(\vmu^*,\,\mSigma^*,\,\vd^*)$
\Comment{The `$*$' superscript indicates final value}
\end{algorithmic}
\end{algorithm}

The MM equations (\ref{eq:mmatch}) do not always admit a solution due to constraint (\ref{eq:constraint}).
Using constraint (\ref{eq:constraint}) and the Woodbury identity, it can be shown that when
$$
(\vs^{\odot1/3})^T\mC^{-1}(\vs^{\odot1/3}) \geq\frac{\sqrt[3]{2}(4-\pi)^{2/3}}{\pi-2},
$$
there is no solution to (\ref{eq:mmatch}).
This intuitively corresponds to cases where the skewness is too large compared to the covariance.
In such cases, one may make an adjustment to the observed value of $\vv=\vs^{\odot1/3}$, of the form $\vv_a=a\vv$, such that a solution does exist.
We can choose $a\in(0,\sqrt{\sqrt[3]{2}(4-\pi)^{2/3}/((\pi-2)\vv^T\mC^{-1}\vv}))$ to minimize the loss function
\begin{align*}
    L(a)=w\norm{\vv_a-\vv}+\norm{\vd_a},
\end{align*}
where $w>0$ is some weight and $\vd_a$ is the matched value of $\vd$ using $\vv_a$ instead of $\vv$.
Intuitively, the left term ensures that the adjustment is not too large (thus compromising the approximation), while the right term ensures that the skewness parameter does not get too large (thus affecting the interpretability of the approximation).
We set $w=2000$ in the simulations; in practice, some manual tuning is required for a given model, where $w$ is chosen so that the matched values of $\vd$ never get too large.

\subsection{Derivative matching}

In the derivative matching (DM) scheme, we denote the mode by $\vm\in\mathbb{R}^p$, the negative Hessian at the mode by $\mJ\in\mathbb{R}^{p\times p}$, and the third-order unmixed derivatives (TUD) at the mode of the observed log-posterior by $\vt\in\mathbb{R}^p$. These are matched to the corresponding values of the MSN density to form an approximation of the posterior distribution $p(\vtheta|\sD)$.
This can be viewed as an extension of the first order Laplace approximation, though \cite{durante2023skewed} formalizes this approach using the GSN distribution.
The final system of matching equations for the derivative matching scheme are given by
\begin{subequations}\label{eq:derivmatch}
\begin{align}
    \vzero&=-\mSigma^{-1}(\vm-\vmu)+\zeta_1(\kappa)\vd, \label{eq:dma} \\
    \mJ&=\mSigma^{-1}-\zeta_2(\kappa)\vd\vd^\top, \label{eq:dmb} \\
    \vt&=\zeta_3(\kappa)\vd^{\odot3},\ \mbox{and} \label{eq:dmc} \\
    \kappa&=\vd^\top(\vm -\vmu), \label{eq:dmd}
\end{align}
\end{subequations}
where $\zeta_k(x)=d^k \log\Phi(x)/dx^k$.
The derivations are provided in the supplementary material.
Here, we have extracted the common term $\kappa$ appearing in (\ref{eq:dma})-(\ref{eq:dmc}) as an additional equation, which facilitates a (unique) solution to (\ref{eq:derivmatch}).
The key idea is to reduce the original system of matching equations to a line search in one dimension for $\kappa$. Once $\kappa$ is solved for, the parameters of the MSN density are then easily recovered.
The steps of the derivative matching scheme are outlined in Algorithm \ref{alg:dm}.

\begin{algorithm}
\caption{Skew-normal matching method: derivative matching}\label{alg:dm}
\begin{algorithmic}[1]
\Require $\vm\in\mathbb{R}^p,\ \mJ\in\mathbb{R}^{p\times p},\ \vt\in\mathbb{R}^p$
\State $\ds R \gets (\vt^{\odot1/3})^\top\mJ^{-1}(\vt^{\odot1/3})$
\State $\ds \kappa^* \gets \text{Solution to }
[\kappa\cdot\zeta_3(\kappa)^{2/3} - R\cdot\zeta_1(\kappa)]\cdot[\zeta_3(\kappa)^{2/3} +R\cdot\zeta_2(\kappa)]  + [R^2\cdot\zeta_1(\kappa)\cdot\zeta_2(\kappa)  ] = 0$  
\State $\ds \vd^* \gets \left[ \vt/\zeta_3(\kappa^*) \right]^{\odot 1/3}$
\State $\ds \mSigma^* \gets \left(\mJ+\zeta_2(\kappa^*)\vd^*(\vd^*)^\top\right)^{-1}$
\State $\ds \vmu^* \gets \vm-\zeta_1(\kappa^*)\mSigma^*\vd^*$ \\
\Return $(\vmu^*,\,\mSigma^*,\,\vd^*)$
\Comment{The `$*$' superscript indicates final value}
\end{algorithmic}
\end{algorithm}

\subsection{Mean-mode-Hessian matching}

In the mean-mode-Hessian (MMH) matching scheme, we use the mode $\vm\in\mathbb{R}^p$, negative Hessian  of the observed log joint likelihood at the mode $\mJ\in\mathbb{R}^{p\times p}$ (which we will assume to be strictly positive definite), and the posterior mean $\widetilde{\vmu}\in\mathbb{R}^p$.
We match these with the corresponding quantities of the MSN density.
This leads to an approximation of the posterior distribution $p(\vtheta|\sD)$.
Note that the difference between the mode and mean provides information about the posterior skewness.
In addition, the mean may be taken from an existing Gaussian approximation, resulting in a post-hoc skewness adjustment.
The system of matching equations for the MMH matching scheme is given by
\begin{subequations}\label{eq:mHmatch}
\begin{align}
    \vzero&=-\mSigma^{-1}(\vm-\vmu)+\zeta_1(\kappa)\vd, \label{eq:mhma} \\
    \mJ&=\mSigma^{-1}-\zeta_2(\kappa)\vd\vd^\top, \label{eq:mhmb} \\
    \widetilde{\vmu}&=\vmu+\sqrt{\frac{2}{\pi}}\cdot\frac{\mSigma\vd}{\sqrt{1+\vd^\top\mSigma\vd}},\ \mbox{and} \label{eq:mhmc} \\
    \kappa&=\vd^\top(\vm -\vmu). \label{eq:mhmd}
\end{align}
\end{subequations}
The derivations are provided in the supplementary material.
Similar to the DM scheme, an auxiliary parameter/equation for $\kappa$ is introduced to facilitate solving (\ref{eq:mHmatch}). The steps of the MMH matching scheme are outlined in Algorithm \ref{alg:mhm}.
Again, the solution reduces the original system of matching equations to a line search in one dimension for $\kappa$ (see line 5 in Algorithm \ref{alg:mhm}). 
Once $\kappa$ is obtained, the parameters of the MSN density are then easily recovered.

\begin{algorithm}
\caption{Skew-normal matching method: mean-mode-Hessian matching}\label{alg:mhm}
\begin{algorithmic}[1]
\Require $\vm\in\mathbb{R}^p,\ \mJ\in\mathbb{R}^{p\times p},\ \widetilde{\vmu}\in\mathbb{R}^p$
\State $\vDelta \gets \widetilde{\vmu}-\vm$
\State $\ds Q \gets \vDelta^\top\mJ\vDelta$
\State $\ds \lambda(\kappa)\coloneqq\sqrt{\frac{2}{\pi}}\cdot\left(1+\frac{\kappa}{\zeta_1(\kappa)}\right)^{-1/2} - \zeta_1(\kappa)$
\State $\ds \alpha(\kappa)\coloneqq\left(\lambda(\kappa)^2/ \zeta_2(\kappa)  - Q - \zeta_2(\kappa) \lambda(\kappa)^2 \left( \frac{\kappa}{\zeta_1(\kappa)} \right)^2\right)^{-1}$
\State $\ds \kappa^* \gets \text{Solution to } \ \kappa\lambda(\kappa)^2 - Q\kappa\lambda(\kappa)^2 \alpha(\kappa) - Q\zeta_1(\kappa)=0$
\State $\ds \mSigma^* \gets \mJ^{-1}-\alpha(\kappa^*)\vDelta\vDelta^\top$
\State $\ds \vd^* \gets \frac{1}{\lambda(\kappa^*)}{(\mSigma^*)^{-1}\vDelta}$
\State $\ds \vmu^* \gets \widetilde{\vmu} - \sqrt{\frac{2}{\pi}} \cdot \frac{\mSigma^*\vd^*}{\sqrt{1+(\vd^*)^\top\mSigma^*\vd^*}}$ \\
\Return $(\vmu^*,\,\mSigma^*,\,\vd^*)$
\Comment{The `$*$' superscript indicates final value}
\end{algorithmic}
\end{algorithm}

Note that $\vDelta$ in Algorithm \ref{alg:mhm} is the difference between the mean and the mode of the target distribution, and indicates the strength and direction of skewness.
It is assumed to be non-zero in the algorithm; the case for $\vDelta=0$ can be reduced to moment matching a multivariate Gaussian and is trivial.

\subsection{Mean-mode-covariance matching}

In the mean-mode-covariance (MMC) matching scheme, the mode $\vm\in\mathbb{R}^p$ of the observed log-posterior, along with the mean $\widetilde{\vmu}\in\mathbb{R}^p$ and covariance matrix $\mC\in\mathcal{S}_+^{p}$ of the posterior $p(\vtheta|\sD)$ are estimated and then matched to the MSN density to form an approximation of the posterior distribution $p(\vtheta|\sD)$.
Care should be taken to avoid confusion with the MM scheme, which also uses the mean and covariance.
Note that the mean and covariance may be taken from an existing Gaussian approximation in practice, resulting in a post-hoc skewness adjustment.
The final system of matching equations for the mean-covariance matching scheme is given by
\begin{subequations}\label{eq:mcmatch}
\begin{align}
    \vzero&=-\mSigma^{-1}(\vm-\vmu)+\zeta_1(\kappa)\vd, \label{eq:mcma} \\
    \widetilde{\vmu}&=\vmu+\sqrt{\frac{2}{\pi}}\cdot\frac{\mSigma\vd}{\sqrt{1+\vd^\top\mSigma\vd}}, \label{eq:mcmb} \\
    \mC&=\mSigma - \frac{2}{\pi(1+\vd^\top\mSigma\vd)}\mSigma\vd\vd^\top\mSigma,\ \mbox{and} \label{eq:mcmc} \\
    \kappa&=\vd^\top(\vm -\vmu). \label{eq:mcmd}
\end{align}
\end{subequations}
The derivations are provided in the supplementary material. Similarly to the DM and MMH schemes, an auxiliary variable $\kappa$ is introduced. The solution again hinges on solving a line search for $\kappa$ after which the parameters of the MSN density are easily recovered. The steps of the mean-mode-covariance matching scheme are outlined in Algorithm \ref{alg:mcm}. Again, note that $\vDelta$ in Algorithm \ref{alg:mcm} indicates the strength and direction of posterior skewness, and is assumed to be non-zero.

\begin{algorithm}
\caption{Skew-normal matching method: Mean-mode-covariance matching}\label{alg:mcm}
\begin{algorithmic}[1]
\Require $\vm\in\mathbb{R}^p,\ \widetilde{\vmu}\in\mathbb{R}^p,\ \mC\in\mathcal{S}_+^{p}$
\State $\ds \vDelta \gets \widetilde{\vmu}-\vm$  
\State $\ds G \gets \vDelta^\top\mC^{-1}\vDelta$ 
\medskip 
\State $\ds \lambda(\kappa)\coloneqq\sqrt{\frac{2}{\pi}}\cdot\left(1+\frac{\kappa}{\zeta_1(\kappa)}\right)^{-1/2} - \zeta_1(\kappa)$
\State $\ds \beta(\kappa)\coloneqq\frac{2}{\pi}\left(1+\frac{\kappa}{\zeta_1(\kappa)}\right)^{-1}\cdot\frac{1}{\lambda(\kappa)^2}$
\State $\ds \kappa^* \gets \text{Solution to }\frac{1}{G}+\beta(\kappa)-\frac{\zeta_1(\kappa)}{\kappa\lambda(\kappa)^2}=0$
\State $\ds \mSigma^* \gets \mC+\beta(\kappa^*)\vDelta\vDelta^\top$
\State $\ds \vd^* \gets  {(\mSigma^*)^{-1}\vDelta}/\lambda(\kappa^*)$
\State $\ds \vmu^* \gets \widetilde{\vmu} - \sqrt{\frac{2}{\pi}}\cdot\frac{\mSigma^*\vd^*}{\sqrt{1+{(\vd^*)^\top\mSigma^*\vd^*}}}$ \\
\Return $(\vmu^*,\,\mSigma^*,\,\vd^*)$
\Comment{The `$*$' superscript indicates final value}
\end{algorithmic}
\end{algorithm}

The MMC equations (\ref{eq:mcmatch}) do not always admit a solution.
When $G\geq2/(\pi-2)$, step 5 in Algorithm \ref{alg:mcm} has no solution (see the supplementary material for more details).
This intuitively corresponds to cases where the skewness is too large compared to the covariance.
In such cases, we make an adjustment to the observed value of $\vDelta$, of the form $\vDelta_a=a\vDelta$, such that a solution does exist.
We can choose $a\in(0,\sqrt{2/[(\pi-2)G]})$ to minimize the loss function
$$
    L(a)=w\norm{\vDelta_a-\vDelta}+\norm{\vd_a},
$$
where $w>0$ is some weight and $\vd_a$ is the matched value of $\vd$ using $\vDelta_a$ instead of $\vDelta$.
The left term ensures that the adjustment is not too large, while the right term ensures that the skewness parameter is not too large.
We set $w=50$ in the simulations but in practice, as with the MM scheme, some manual tuning is required for a given model.

\section{Simulation and benchmark settings}\label{sec:settings}

The following sections describe practical applications of the skew-normal matching method for simple Bayesian problems, where its performance was measured on both simulated and benchmark datasets.
In particular, the case of Bayesian probit regression was considered, with logistic regression results provided in the supplementary material.
The purpose of this section is to outline the settings that were used throughout these simulations and benchmarks.

\subsection{Additional methods compared}

A selection of common ABI methods were chosen to be compared to the skew-normal matching method.
The first of these is Laplace's method, which has seen widespread use in the approximation of posteriors and posterior integrals \citep{tierney_kadane_1986, raudenbush_yang_yosef_2000, rue_martino_chopin_2009}.
Next, we considered Gaussian variational Bayes (GVB), another well-established method, with direct optimization of the evidence lower bound performed via the Broyden–Fletcher–Goldfarb–Shanno (BFGS) algorithm.
As a related method with relative ease of implementation, the delta method variational Bayes (DMVB) variant of \cite{braun2010variational} and \cite{wang2013variational} was also chosen.
The expectation propagation (EP) framework from \cite{minka_2001}, in its classical Gaussian product-over-sites implementation, was also investigated.
Finally, the SM approximation of \cite{durante2023skewed} (with 50,000 samples used for posterior inference) was considered as a potential theoretically justified alternative to the skew-normal matching method.

\subsection{Simulation settings}

All methods considered were subject to the same simulation settings for the purpose of comparison.
For each combination of method, $p\in\left\{2,4,8,16,24,40\right\}$, and $n\in\left\{2p,4p\right\}$, either random Gaussian ($X_{ij}\sim\text{i.i.d.}\ \mathcal{N}_1(0,1)$ for $i=1,\ldots,n$ and $j=2,\ldots,p$) or AR1 covariate data ($\left\{X_{i2},\ldots,X_{ip}\right\}\sim\text{AR}(1)$ with coefficient $\rho=0.9$, for $i=1,\ldots,n$) was generated.
In both cases, $X_{i1}=1$, for $i=1,\ldots,n$.
Call $n=2p$ the low data case and $n=4p$ the moderate data case; we chose $n$ to be small relative to $p$ so as to induce more difficult situations where the posterior distribution can become quite skewed.
The corresponding response data was then randomly generated using $\vtheta_p=(2,-2,\cdots,2,-2)^\top/p$, with $\vtheta_p$ having $p$ entries.
Data was simulated 50 times for each combination of settings; in all cases, the Gaussian prior was centered at zero, with the variance $\sigma_\vtheta^2$ set to $10,000$.
Finally, in order to avoid pathological cases, we discarded simulations where separation was detected in the data (see \cite{mansournia2018separation} for an overview).

\subsection{Benchmark datasets}

In addition to the simulated examples above, a selection of eight commonly used benchmark datasets for binary classification from the UCI machine learning repository were also used to evaluate the performance of the skew-normal matching method.
Each dataset contained a certain number of numeric and categorical predictors, in addition to a column of binary responses to be predicted on.
These datasets were coded as O-rings ($n=23$, $p=3$), Liver ($n=345$, $p=6$), Diabetes ($n=392$, $p=9$), Glass ($n=214$, $p=10$), Breast cancer ($n=699$, $p=10$), Heart ($n=270$, $p=14$), German credit ($n=1000$, $p=25$), and Ionosphere ($n=351$, $p=33$), where the value $p$ includes the intercept.
Breast cancer, German credit, and Ionosphere were shortened to Breast, German, and Ion respectively.

\subsection{Performance evaluation}

In each simulation or instance of benchmark data, the corresponding posterior distribution was initially computed using the No-U-Turn sampler, as implemented via the R package \texttt{rstan} \citep{rlang,rstan}.
Each posterior was approximated using a total of 50,000 MCMC iterations, 5,000 of which were warm-up iterations.
Convergence was verified by checking that the $\hat{R}$ value was less than $1.1$ \citep{Gelman1995}.
This MCMC approximation, if the chain converged, acted as the gold standard to which the rest of the methods were compared.
If the chain did not converge, the result was discarded.
For each method, comparisons were made across marginal components -- the difference between the marginal components of the approximation and the corresponding marginal components of the MCMC estimate were computed using the $L^1$ norm.
An appropriate transformation was then performed on this norm (ranging between 0 and 2) in order to obtain the $L^1$ accuracy.
The $L^1$ accuracy for the $j$-th marginal component is given by
\begin{align}\label{eq:l1acc}
    \text{$L^1$ accuracy}&=1-\frac{1}{2}\int_{-\infty}^\infty\left|p_\text{MCMC}(\theta_j)-p_\text{Approx.}(\theta_j)\right|\,d\theta_j
\end{align}
where $p_\text{MCMC}(\theta_j)$ and $p_\text{Approx.}(\theta_j)$ are the $j$-th marginal components of the MCMC and approximation, respectively.
The former was estimated via \texttt{density()}, the default kernel density estimation function in R.
The $L^1$ accuracy ranges in value from 0 to 1, with values closer to one indicating better performance.
The integral in (\ref{eq:l1acc}) was computed numerically, with the bounds of integration set to $m_j\pm5\sqrt{\mSigma_{j,j}}$, with $\vm$ and $\mSigma$ being the mean and covariance respectively of the MCMC samples.
This range was evenly split into 1,000 intervals for use with the composite trapezoidal rule.

\section{Applications as a standalone approximation}\label{sec:standalone}

In this section, the skew-normal matching method was applied as a standalone approximation to Bayesian probit regression and compared to other methods.
We use the word standalone in the sense that the posterior statistics required in the skew-normal matching algorithm are estimated using an accurate, non-Gaussian-based approach.
For moment matching, $\widetilde{\vmu}$, $\mC$, and $\vs$ were all estimated using Pareto smoothed importance sampling (PSIS) \citep{vehtari2021}.
For derivative matching, $\vm$, $\mJ$, and $\vt$ were all calculated using the Newton-Raphson method.
For MMH matching, $\vm$ and $\mJ$ were estimated using Newton-Raphson, while $\widetilde{\vmu}$ was estimated using one of three methods: a Jensen-based approach, an improved Laplace approach, and PSIS (denoted MMH-Jensen, MMH-IL, and MMH-IS, respectively).
Finally, for MMC matching, $\vm$ was estimated via Newton-Raphson, while $\widetilde{\vmu}$ and $\mC$ were estimated using PSIS.
Note that all matching methods costs $O(p^2)$ once these posterior statistics are calculated and are not in the form of an iterative solution so no convergence checks are required.
The main costs are often in calculating the posterior statistics themselves (this can be avoided by only considering post-hoc adjustments, and is the focus of the next section).
Derivations for the estimation of posterior statistics and the Jensen variant of MMH matching, in addition to plots, are given in the supplementary material.

\subsection{Simulation results}

\begin{table}
\centering
\resizebox{\linewidth}{!}{
\begin{tabular}{ccccccccccccc}
\toprule
\multicolumn{1}{c}{ } & \multicolumn{2}{c}{$p=2$} & \multicolumn{2}{c}{$p=4$} & \multicolumn{2}{c}{$p=8$} & \multicolumn{2}{c}{$p=16$} & \multicolumn{2}{c}{$p=24$} & \multicolumn{2}{c}{$p=40$} \\
\cmidrule(l{3pt}r{3pt}){2-3} \cmidrule(l{3pt}r{3pt}){4-5} \cmidrule(l{3pt}r{3pt}){6-7} \cmidrule(l{3pt}r{3pt}){8-9} \cmidrule(l{3pt}r{3pt}){10-11} \cmidrule(l{3pt}r{3pt}){12-13}
Method & Acc. & Time & Acc. & Time & Acc. & Time & Acc. & Time & Acc. & Time & Acc. & Time\\
\midrule
\addlinespace[0.3em]
\multicolumn{13}{l}{\textbf{$\mathbf{n=2p}$}}\\
\hspace{1em}Laplace & 76.7 & 0.01 & 75.0 & 0.00 & 72.1 & 0.00 & 68.8 & 0.00 & 68.3 & 0.00 & 62.2 & 0.01\\
\hspace{1em}DMVB & 75.9 & 0.01 & 71.9 & 0.01 & 64.5 & 0.04 & 64.2 & 0.19 & 63.1 & 0.52 & 58.7 & 3.01\\
\hspace{1em}GVB & 86.2 & 0.04 & 86.3 & 0.07 & 88.6 & 0.21 & 89.0 & 0.83 & 90.6 & 1.50 & 88.7 & 2.91\\
\hspace{1em}EP & 86.0 & 0.01 & 89.5 & 0.00 & 92.8 & 0.01 & \textbf{94.3} & 0.01 & \textbf{95.8} & 0.04 & \textbf{96.0} & 0.10\\
\hspace{1em}SM & 78.7 & 0.53 & 80.7 & 0.88 & 78.3 & 1.56 & 74.8 & 2.94 & 74.2 & 4.37 & 67.3 & 7.68\\
\hspace{1em}MM & 96.2 & 2.12 & 94.9 & 2.78 & 93.7 & 3.00 & 92.1 & 3.42 & 90.4 & 3.83 & 83.0 & 4.98\\
\hspace{1em}DM & 83.7 & 0.04 & 79.7 & 0.06 & 78.1 & 0.12 & 73.5 & 0.26 & 76.5 & 0.37 & 69.1 & 0.62\\
\hspace{1em}MMH-Jensen & 90.9 & 0.04 & 89.7 & 0.08 & 88.2 & 0.16 & 85.5 & 0.38 & 85.9 & 0.60 & 80.7 & 1.52\\
\hspace{1em}MMH-IL & 92.4 & 0.07 & 91.7 & 0.19 & 91.4 & 0.50 & 90.5 & 1.36 & 90.5 & 2.76 & 86.5 & 11.09\\
\hspace{1em}MMH-IS & 94.3 & 2.12 & 93.6 & 2.78 & 93.3 & 2.99 & 91.7 & 3.43 & 90.1 & 3.85 & 85.6 & 4.99\\
\hspace{1em}MMC & \textbf{96.5} & 2.11 & \textbf{95.0} & 2.77 & \textbf{94.8} & 3.00 & 92.6 & 3.44 & 90.5 & 3.86 & 83.0 & 5.03\\
\addlinespace[0.3em]
\multicolumn{13}{l}{\textbf{$\mathbf{n=4p}$}}\\
\hspace{1em}Laplace & 78.6 & 0.00 & 83.3 & 0.00 & 86.2 & 0.00 & 90.3 & 0.00 & 91.4 & 0.00 & 92.7 & 0.01\\
\hspace{1em}DMVB & 78.4 & 0.01 & 85.9 & 0.01 & 92.1 & 0.03 & 97.1 & 0.19 & 98.2 & 0.67 & 98.7 & 3.92\\
\hspace{1em}GVB & 85.8 & 0.05 & 91.7 & 0.10 & 95.5 & 0.31 & 98.1 & 0.84 & 98.7 & 1.24 & 98.9 & 3.35\\
\hspace{1em}EP & 85.6 & 0.01 & 93.2 & 0.01 & 96.4 & 0.01 & 98.5 & 0.02 & \textbf{98.9} & 0.05 & \textbf{99.2} & 0.13\\
\hspace{1em}SM & 80.9 & 0.53 & 90.5 & 0.86 & 92.6 & 1.54 & 94.9 & 2.88 & 95.6 & 4.32 & 96.0 & 7.68\\
\hspace{1em}MM & \textbf{96.9} & 2.11 & \textbf{97.5} & 2.72 & \textbf{98.0} & 2.97 & 98.4 & 3.43 & 98.5 & 3.93 & 98.3 & 5.23\\
\hspace{1em}DM & 86.9 & 0.04 & 87.9 & 0.06 & 88.7 & 0.14 & 93.3 & 0.26 & 93.9 & 0.38 & 94.8 & 0.63\\
\hspace{1em}MMH-Jensen & 93.9 & 0.05 & 95.6 & 0.08 & 96.6 & 0.18 & 97.8 & 0.43 & 98.2 & 0.82 & 98.3 & 2.55\\
\hspace{1em}MMH-IL & 94.8 & 0.08 & 95.9 & 0.21 & 96.8 & 0.51 & 97.6 & 1.48 & 97.9 & 3.39 & 98.1 & 14.33\\
\hspace{1em}MMH-IS & 95.9 & 2.11 & 96.2 & 2.72 & 96.9 & 2.97 & 97.8 & 3.44 & 98.0 & 3.94 & 98.0 & 5.25\\
\hspace{1em}MMC & 96.7 & 2.11 & 97.3 & 2.72 & \textbf{98.0} & 2.97 & \textbf{98.6} & 3.44 & 98.6 & 3.94 & 98.4 & 5.26\\
\bottomrule
\end{tabular}}
\caption{Mean marginal $L^1$ accuracies (Acc.) (expressed as a percentage) and times (in seconds) across probit regression simulations with independent covariate data. The highest accuracies for each value of $p$ appear in bold.}
\label{tab:probitsims}
\end{table}

\begin{figure}
    \begin{center}
        \includegraphics[scale=0.65]{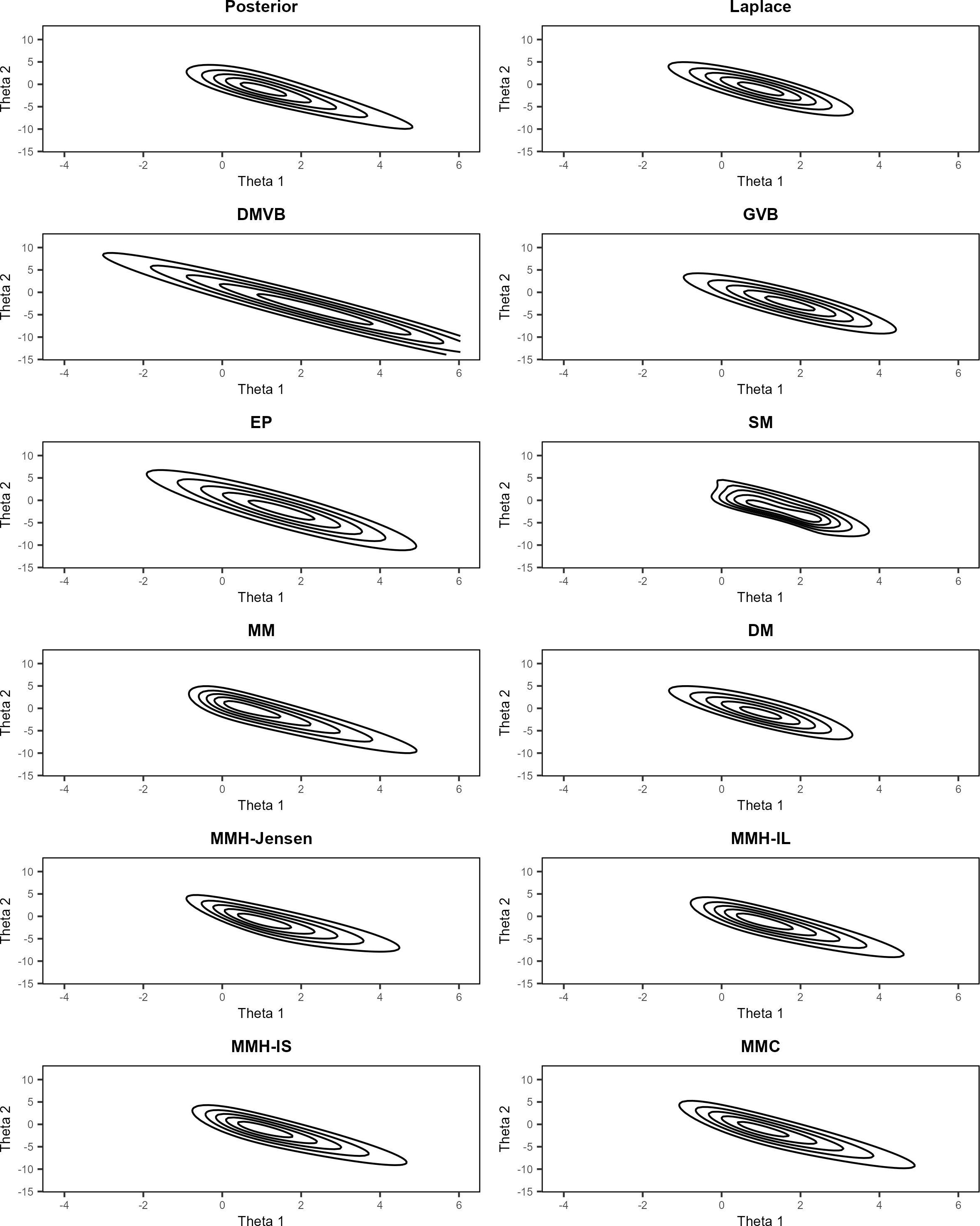}
    \end{center}
    \caption{Contour plots for the posterior and across various approximation methods, for a small probit regression example with $p=2$ and $n=4$.}
    \label{pr-contours}
\end{figure}

Simulation results for probit regression with independent covariates are shown in Table \ref{tab:probitsims}. The other cases are included in the supplementary material.
Moment matching and mean-mode-covariance matching tended to have the highest $L^1$ accuracies, followed by mean-mode-Hessian matching and, finally, by derivative matching.
The SM approximation had around the same $L^1$ accuracies as derivative matching, with this accuracy improving when both $p$ and $n$ were high.

In both the low and moderate data cases, skew-normal matching approximations generally performed better compared to standard ABI approaches.
However, as $p$ increased, this improvement in accuracy diminished.
Furthermore, as $p$ increased, exact solutions to the moment matching and mean-mode-covariance matching schemes became less common, and, in addition, matching became less effective compared to variational approaches; this also led to decreases in accuracy.
When $p$ is above or equal to 24, GVB and EP started to outperform the skew-normal matching method.

In most cases, the skew-normal matching variants tended to be slower than their Gaussian approximations by an order of around 2 to 100, depending on the setting, with additionally the SM approximation taking roughly the same amount of time as the skew-normal matching variants.
It should be noted that some parts of GVB and SM were implemented in \CC\ for speed, and so a direct comparison with the skew-normal matching method is not available.
GVB is expected to be slower than some of the skew-normal matching variants under a fair comparison.

In general, the variants of the skew-normal matching method are seen to provide an increase in posterior fit compared to Gaussian approaches when $p$ is not too high, at the cost of decreased speed.
Better overall performance for the MM and MMC schemes may be achieved by only performing such an approximation when an exact solution exists, and using a Gaussian approximation otherwise.
Figure \ref{pr-contours} illustrates the benefits of the skew-normal matching method in a particular low-dimensional case with $p=2$ and $n=4$.
Note that, for this specific example, approximate solutions were not needed for the MM and MMC schemes.

\subsection{Benchmark results}

\begin{table}
\centering
\resizebox{\linewidth}{!}{
\begin{tabular}{ccccccccccccccccc}
\toprule
\multicolumn{1}{c}{ } & \multicolumn{2}{c}{O-rings} & \multicolumn{2}{c}{Liver} & \multicolumn{2}{c}{Diabetes} & \multicolumn{2}{c}{Glass} & \multicolumn{2}{c}{Breast} & \multicolumn{2}{c}{Heart} & \multicolumn{2}{c}{German} & \multicolumn{2}{c}{Ion.} \\
\cmidrule(l{3pt}r{3pt}){2-3} \cmidrule(l{3pt}r{3pt}){4-5} \cmidrule(l{3pt}r{3pt}){6-7} \cmidrule(l{3pt}r{3pt}){8-9} \cmidrule(l{3pt}r{3pt}){10-11} \cmidrule(l{3pt}r{3pt}){12-13} \cmidrule(l{3pt}r{3pt}){14-15} \cmidrule(l{3pt}r{3pt}){16-17}
Method & Acc. & Time & Acc. & Time & Acc. & Time & Acc. & Time & Acc. & Time & Acc. & Time & Acc. & Time & Acc. & Time\\
\midrule
Laplace & 92.5 & 0.17 & 98.0 & 0.00 & 98.3 & 0.00 & 92.4 & 0.00 & 95.3 & 0.05 & 95.7 & 0.00 & 97.9 & 0.11 & 86.0 & 0.03\\
DMVB & 95.5 & 0.14 & 99.1 & 0.19 & \textbf{99.2} & 0.61 & 96.9 & 0.33 & 97.1 & 1.97 & \textbf{99.1} & 0.74 & \textbf{99.2} & 31.00 & 93.2 & 10.85\\
GVB & 96.5 & 0.33 & 99.1 & 0.09 & \textbf{99.2} & 0.06 & 97.3 & 2.16 & 98.2 & 2.12 & 99.0 & 0.09 & \textbf{99.2} & 0.46 & 97.3 & 11.54\\
EP & 96.6 & 0.32 & 99.1 & 0.14 & \textbf{99.2} & 0.14 & 97.8 & 0.08 & 98.4 & 0.30 & \textbf{99.1} & 0.11 & \textbf{99.2} & 0.67 & \textbf{98.6} & 0.35\\
SM & 96.9 & 0.91 & 99.2 & 1.14 & 99.0 & 1.62 & 96.2 & 2.03 & 97.9 & 1.81 & 98.5 & 2.61 & 99.0 & 4.47 & 90.5 & 6.05\\
\addlinespace
MM & \textbf{99.1} & 2.45 & 99.1 & 3.44 & 99.1 & 4.07 & \textbf{98.2} & 3.39 & \textbf{98.6} & 5.39 & 98.9 & 3.63 & 99.0 & 7.70 & 96.5 & 6.05\\
DM & 95.7 & 0.32 & 98.3 & 0.12 & 98.4 & 0.17 & 52.9 & 0.18 & 96.2 & 0.20 & 97.1 & 0.25 & 95.3 & 0.47 & 86.4 & 0.51\\
MMH-Jensen & 98.5 & 0.36 & \textbf{99.3} & 0.37 & \textbf{99.2} & 0.93 & 97.3 & 0.42 & 98.4 & 2.20 & \textbf{99.1} & 1.06 & 99.1 & 21.46 & 96.0 & 5.75\\
MMH-IL & 98.6 & 0.53 & \textbf{99.3} & 2.19 & \textbf{99.2} & 6.01 & 97.5 & 3.64 & 98.4 & 21.26 & 99.0 & 4.90 & 99.1 & 125.89 & 97.1 & 39.86\\
MMH-IS & 98.5 & 2.45 & 99.2 & 3.45 & \textbf{99.2} & 4.06 & 97.6 & 3.31 & 98.4 & 5.42 & 98.9 & 3.70 & 99.0 & 7.69 & 96.7 & 6.04\\
\addlinespace
MMC & 98.8 & 2.42 & 99.2 & 3.46 & \textbf{99.2} & 4.03 & 98.1 & 3.33 & \textbf{98.6} & 5.38 & 98.9 & 3.67 & 99.0 & 7.81 & 96.5 & 6.05\\
\bottomrule
\end{tabular}}
\caption{Mean marginal $L^1$ accuracies (Acc.) (expressed as a percentage) across probit regression benchmarks. The highest accuracies for each benchmark dataset appear in bold.}
\label{tab:probitbench}
\end{table}

Results for the benchmark datasets are presented in Table \ref{tab:probitbench}.
A similar hierarchy of performance to the simulations was observed among the skew-normal matching variants, with moment matching, mean-mode-covariance, and mean-mode-Hessian matching generally performing better, and derivative matching performing worse.
Here, the SM approximation had higher $L^1$ accuracies compared to derivative matching, but lower accuracies compared to the other skew-normal matching variants.
The skew-normal matching method performed as well or better compared to its Gaussian approximation methods when $p$ was not too high, as was the case for the German credit and Ionosphere datasets.
For Ionosphere in particular, approximate solutions were required for the moment matching and mean-mode-covariance matching schemes.
As with before, the skew-normal matching method comes with the cost of decreased speed.

\section{Applications as a post-hoc adjustment}\label{sec:post-hoc}

The use of importance sampling to estimate key statistics in the moment matching and mean-mode-covariance matching schemes in Section \ref{sec:standalone} can be a deterrent due to the non-negligible cost of needing to use importance sampling, especially for larger problems.
We believe that the practicality of the skew-normal matching method lies mainly in post-hoc skewness adjustments to standard Gaussian approximations.
That is, the required posterior statistics are estimated by the standard Gaussian approximations themselves.
The original Gaussian approximation are then adjusted post-hoc, by using the skew-normal matching method.
The costs of the post-hoc adjustments are minimal.

In this section we investigate the performance of two types of post-hoc adjustments.
The first is a mean-mode-Hessian adjustment.
Here, an approximate posterior mean $\vmu$ is first taken from a given Gaussian base approximation (we use one of DMVB, GVB or EP in this paper, as described in Section \ref{sec:settings}).
The mode $\vm$ and negative Hessian at the mode $\mJ$ of the posterior are then calculated via Newton-Raphson.
Finally, these three quantities are used as inputs to Algorithm \ref{alg:mhm} to produce a skew-normal approximation.
This approximation can be considered as a post-hoc skewness adjustment to the original Gaussian base approximation.

The second is a mean-mode-covariance adjustment.
Here, an approximate posterior mean $\vmu$ and covariance $\mSigma$ are first taken from a given Gaussian base approximation (again, one of DMVB, GVB or EP).
The mode $\vm$ of the posterior is then calculated via Newton-Raphson.
Finally, these three quantities are used as inputs to Algorithm \ref{alg:mcm} to produce a skew-normal approximation.
This approximation can again be considered as a post-hoc skewness adjustment to the original Gaussian base approximation.

Approximate solutions were not used for the mean-covariance adjustment when no solution was obtained from either Algorithm \ref{alg:mhm} or Algorithm \ref{alg:mcm}, with the justification being that there is a Gaussian approximation to fall back to.
In such cases, the result was discarded.
As with before, both simulations and benchmark datasets (under the same settings from Section \ref{sec:settings}) were used to measure performance. 
Plots can be found in the supplementary material.

\subsection{Simulation results}

\begin{table}
\centering
\resizebox{\linewidth}{!}{
\begin{tabular}{ccccccccccccc}
\toprule
\multicolumn{1}{c}{ } & \multicolumn{2}{c}{$p=2$} & \multicolumn{2}{c}{$p=4$} & \multicolumn{2}{c}{$p=8$} & \multicolumn{2}{c}{$p=16$} & \multicolumn{2}{c}{$p=24$} & \multicolumn{2}{c}{$p=40$} \\
\cmidrule(l{3pt}r{3pt}){2-3} \cmidrule(l{3pt}r{3pt}){4-5} \cmidrule(l{3pt}r{3pt}){6-7} \cmidrule(l{3pt}r{3pt}){8-9} \cmidrule(l{3pt}r{3pt}){10-11} \cmidrule(l{3pt}r{3pt}){12-13}
Adjust. & $+$ & $-$ & $+$ & $-$ & $+$ & $-$ & $+$ & $-$ & $+$ & $-$ & $+$ & $-$\\
\midrule
\addlinespace[0.3em]
\multicolumn{13}{l}{\textbf{$\mathbf{n=2p}$}}\\
\addlinespace[0.3em]
\multicolumn{13}{l}{\textbf{DMVB}}\\
\hspace{1em}\hspace{1em}MMH & \textbf{22.1} & \textbf{1.5} & \textbf{57.7} & \textbf{1.2} & \textbf{153.4} & \textbf{0.9} & \textbf{271.3} & \textbf{3.7} & \textbf{395.9} & \textbf{5.9} & \textbf{644.1} & \textbf{16.1}\\
\hspace{1em}\hspace{1em}MMC & \textbf{3.2} & \textbf{0.4} & \textbf{5.4} & \textbf{0.1} & \textbf{4.5} & \textbf{0.0} & \textbf{7.0} & \textbf{0.0} & \textbf{4.9} & \textbf{0.0} & \text{---} & \text{---}\\
\addlinespace[0.3em]
\multicolumn{13}{l}{\textbf{GVB}}\\
\hspace{1em}\hspace{1em}MMH & \textbf{16.6} & \textbf{0.7} & \textbf{33.4} & \textbf{3.1} & \textbf{48.6} & \textbf{7.3} & \textbf{76.3} & \textbf{14.6} & \textbf{74.6} & \textbf{26.4} & \textbf{104.4} & \textbf{66.1}\\
\hspace{1em}\hspace{1em}MMC & \textbf{3.8} & \textbf{0.1} & \textbf{3.2} & \textbf{0.3} & \textbf{2.8} & \textbf{0.2} & \textbf{1.5} & \textbf{0.2} & \text{---} & \text{---} & \text{---} & \text{---}\\
\addlinespace[0.3em]
\multicolumn{13}{l}{\textbf{EP}}\\
\hspace{1em}\hspace{1em}MMH & \textbf{7.8} & \textbf{0.5} & \textbf{12.5} & \textbf{5.6} & \textbf{21.7} & \textbf{17.4} & 27.1 & 40.4 & 15.6 & 82.1 & 11.5 & 209.6\\
\hspace{1em}\hspace{1em}MMC & \textbf{4.2} & \textbf{0.4} & \textbf{7.0} & \textbf{0.3} & \textbf{8.4} & \textbf{0.0} & \textbf{7.2} & \textbf{0.0} & \text{---} & \text{---} & \text{---} & \text{---}\\
\addlinespace[0.3em]
\multicolumn{13}{l}{\textbf{$\mathbf{n=4p}$}}\\
\addlinespace[0.3em]
\multicolumn{13}{l}{\textbf{DMVB}}\\
\hspace{1em}\hspace{1em}MMH & \textbf{25.9} & \textbf{0.5} & \textbf{31.2} & \textbf{0.3} & \textbf{30.1} & \textbf{1.5} & \textbf{14.1} & \textbf{7.0} & 7.1 & 13.2 & 4.2 & 26.3\\
\hspace{1em}\hspace{1em}MMC & \textbf{4.7} & \textbf{0.2} & \textbf{7.3} & \textbf{0.0} & \textbf{8.0} & \textbf{0.0} & \textbf{7.0} & \textbf{0.0} & \textbf{6.1} & \textbf{0.0} & \textbf{4.3} & \textbf{0.0}\\
\addlinespace[0.3em]
\multicolumn{13}{l}{\textbf{GVB}}\\
\hspace{1em}\hspace{1em}MMH & \textbf{20.2} & \textbf{0.1} & \textbf{19.6} & \textbf{1.4} & \textbf{14.6} & \textbf{2.7} & 5.6 & 9.6 & 3.8 & 15.6 & 2.5 & 27.9\\
\hspace{1em}\hspace{1em}MMC & \textbf{3.6} & \textbf{0.2} & \textbf{4.2} & \textbf{0.0} & \textbf{3.8} & \textbf{0.2} & \textbf{2.8} & \textbf{0.2} & \textbf{2.1} & \textbf{0.3} & \textbf{1.1} & \textbf{0.3}\\
\addlinespace[0.3em]
\multicolumn{13}{l}{\textbf{EP}}\\
\hspace{1em}\hspace{1em}MMH & \textbf{11.7} & \textbf{0.2} & \textbf{12.3} & \textbf{2.1} & \textbf{9.2} & \textbf{4.0} & 3.7 & 13.4 & 2.4 & 20.6 & 1.2 & 35.5\\
\hspace{1em}\hspace{1em}MMC & \textbf{2.7} & \textbf{1.1} & \textbf{5.4} & \textbf{0.4} & \textbf{5.7} & \textbf{0.1} & \textbf{4.6} & \textbf{0.1} & \textbf{3.6} & \textbf{0.1} & \textbf{2.0} & \textbf{0.1}\\
\bottomrule
\end{tabular}}
\caption{Mean total marginal $L^1$ improvement/deterioration of accuracies (expressed as a percentage) across probit regression post-hoc simulations with independent covariate data. The column `$+$' indicates a mean total improvement in marginal $L^1$ accuracy, while the column `$-$' indicates the mean total decrease in marginal $L^1$ accuracy across simulations. For example, if the marginal improvements in $L^1$ accuracy for a $p=4$ example were $(5, -2, 1, -3)$, then the total marginal $L^1$ improvement and deterioration are 6 and 5 respectively. Pairs where there is an overall improvement are highlighted in bold. Long dashes indicate no data.}
\label{tab:probitphsims}
\end{table}

\begin{table}
\centering
\begin{tabular}{ccccccc}
\toprule
Multiplier & \text{$p=2$} & \text{$p=4$} & \text{$p=8$} & \text{$p=16$} & \text{$p=24$} & \text{$p=40$}\\
\midrule
2 & 92.0 & 79.3 & 48.7 & 17.3 & 0.7 & 0.0\\
4 & 93.3 & 94.0 & 93.3 & 97.3 & 96.7 & 96.7\\
\bottomrule
\end{tabular}
\caption{Success rates, i.e., rates where a solution exists to Algorithm \ref{alg:mcm}, (expressed as a percentage) averaged over base approximations for the mean-mode-covariance based skewness adjustment. The mean-mode-Hessian based adjustments had a 100\% success rate.}
\label{tab:probitph2success}
\end{table}

Simulation results for probit regression with independent covariates are shown in Tables \ref{tab:probitphsims} and \ref{tab:probitph2success}.
We have included the probit case because it showed the most promising results. The other cases are included in the supplementary material.

In the low data case, a mean-mode-Hessian adjustment provided a noticeable increase in the quality of the posterior fit for the DMVB and GVB approximations, while a decrease in the quality of fit was seen for EP.
On the other hand, mean-mode-covariance adjustments tended to increase or maintain $L^1$ accuracy for low values of $p$, with diminishing returns as $p$ increased and exact solutions became rarer (as indicated by Table \ref{tab:probitph2success}).
In the moderate data case, the mean-mode-Hessian adjustment was seen to increase $L^1$ accuracies across all methods for low values of $p$, but slightly decreased these accuracies for GVB and EP as $p$ approached $40$.
In contrast, the mean-mode-covariance adjustment always increased or maintained the quality of the posterior fit.
In general, simulations show that in most cases, some form of post-hoc skewness adjustment works reasonably well if the number of dimensions is not too high.
The mean-Hessian adjustment gave very favorable results when there was low data (excluding EP, where it appeared to break down), while the mean-covariance adjustment gave favorable results in all situations.

\subsection{Benchmark results}

\begin{table}
\centering
\resizebox{\linewidth}{!}{
\begin{tabular}{ccccccccccccccccc}
\toprule
\multicolumn{1}{c}{ } & \multicolumn{2}{c}{O-rings} & \multicolumn{2}{c}{Liver} & \multicolumn{2}{c}{Diabetes} & \multicolumn{2}{c}{Glass} & \multicolumn{2}{c}{Breast} & \multicolumn{2}{c}{Heart} & \multicolumn{2}{c}{German} & \multicolumn{2}{c}{Ion.} \\
\cmidrule(l{3pt}r{3pt}){2-3} \cmidrule(l{3pt}r{3pt}){4-5} \cmidrule(l{3pt}r{3pt}){6-7} \cmidrule(l{3pt}r{3pt}){8-9} \cmidrule(l{3pt}r{3pt}){10-11} \cmidrule(l{3pt}r{3pt}){12-13} \cmidrule(l{3pt}r{3pt}){14-15} \cmidrule(l{3pt}r{3pt}){16-17}
\text{Adjust.} & \text{+} & \text{--} & \text{+} & \text{--} & \text{+} & \text{--} & \text{+} & \text{--} & \text{+} & \text{--} & \text{+} & \text{--} & \text{+} & \text{--} & \text{+} & \text{--}\\
\midrule
\addlinespace[0.3em]
\multicolumn{17}{l}{\textbf{DMVB}}\\
\hspace{1em}MMH & {\bf 7.4} & {\bf 0.4} & {\bf 1.1} & {\bf 0.0} & {\bf 0.6} & {\bf 0.2} & {\bf 6.8} & {\bf 0.5} & {\hspace{-0.25cm}\bf 11.4} & {\bf 0.2} & 1.1       & 1.6       & 1.0 & 2.8 & {\bf 89.2} & {\bf \ 3.4} \\
\hspace{1em}MMC & {\bf 7.1} & {\bf 0.0} & {\bf 1.2} & {\bf 0.0} & {\bf 0.5} & {\bf 0.0} & 0.4       & 0.6  & {\bf 0.9}  & {\bf 0.0}       & {\bf 1.6} & {\bf 0.0} & 0.8 & 1.8 & \text{---} & \text{---} \\
\addlinespace[0.3em]
\multicolumn{17}{l}{\textbf{GVB}}\\
\hspace{1em}MMH & {\bf 6.3} & {\bf 0.4} & {\bf 1.3} & {\bf 0.0} & {\bf 0.7} & {\bf 0.2} & {\bf 4.6} & {\bf 2.0} & {\bf 1.9} & {\bf 0.3} & {\bf 1.2} & {\bf 0.8} & 1.0 & 3.4 & {\bf 24.1} & {\bf 14.1} \\
\hspace{1em}MMC & {\bf 3.4} & {\bf 0.0} & {\bf 1.1} & {\bf 0.0} & {\bf 0.4} & {\bf 0.0} & {\bf 0.5} & {\bf 0.0} & {\bf 0.9} & {\bf 0.0} & {\bf 1.0} & {\bf 0.0} & 0.8 & 1.8 & \text{---} & \text{---}\\
\addlinespace[0.3em]
\multicolumn{17}{l}{\textbf{EP}}\\
\hspace{1em}MMH & {\bf 6.4} & {\bf 0.4} & {\bf 1.3} & {\bf 0.0} & {\bf 0.7} & {\bf 0.3} & 1.8 & 3.4 & {\bf 1.4} & {\bf 1.2} & 1.0 & 1.0 & 1.0 & 3.0 & 1.7 & 34.0 \\
\hspace{1em}MMC & {\bf 4.1} & {\bf 0.0} & {\bf 1.2} & {\bf 0.0} & {\bf 0.4} & {\bf 0.0} & {\bf 0.5} & {\bf 0.4} & {\bf 0.9} & {\bf 0.0} & {\bf 1.0} & {\bf 0.0} & 0.8 & 1.9 & \text{---} & \text{---}\\
\bottomrule
\end{tabular}}
\caption{Mean total marginal improvement/deterioration in $L^1$ accuracies (expressed as a percentage) across probit regression post-hoc benchmarks. The `$+$' columns indicates the total increases in marginal $L^1$ accuracy, while the `$-$' column indicates the total decrease in marginal $L^1$ accuracy. For example, if the marginal improvements in $L^1$ accuracy for a $p=4$ example were $(5, -2, 1, -3)$, then the total marginal $L^1$ improvement and deterioration are 6 and 5 respectively. Pairs where there is an overall improvement are highlighted in bold. Long dashes indicate no data.}
\label{tab:probitphbench}
\end{table}

Results for the benchmarks are shown in Table \ref{tab:probitphbench}.
In the majority of cases, a post-hoc skewness adjustment was seen to give a slight improvement to the $L^1$ accuracy of the original Gaussian approximation.
As with the simulations, mean-mode-Hessian adjustments gave larger increases in $L^1$ accuracy but also had the risk of decreasing accuracy when $p$ was large.
This was seen with the Heart, German credit, and Ionosphere datasets.
On the other hand, mean-covariance adjustments tended to give more modest increases in accuracy but had a much lower chance of decreasing posterior fit.
The German credit dataset appears to be an anomaly in this regard, and requires further investigation.
In general, the benchmarks show that it is beneficial to apply some form of post-hoc skewness adjustment to an existing Gaussian approximation when the number of dimensions is not too high.
Some approximations benefit from a skewness adjustment more than others.

\section{Summary and future work}\label{sec:summary}

This paper has introduced the skew-normal matching method as an alternative to standard Gaussian-based approximations of the posterior distribution, providing a potentially more practical alternative to the SM approximation of \cite{durante2023skewed}.
Four matching schemes are suggested, namely moment matching, derivative matching, mean-mode-Hessian matching, and mean-mode-covariance matching.
Each scheme was based on matching a combination of derivatives at the mode and moments of the posterior with that of the MSN density.
The performance of these matching schemes were evaluated on a Bayesian probit regression model, where it was seen that for both simulated and benchmark data, the skew-normal matching method tended to outperform standard Gaussian-based approximations in terms of accuracy for low and moderate dimensions, at the cost of run time.
All matching schemes except DM were also seen to mostly outperform the SM approximation in the settings investigated.
Each matching scheme is seen to offer a trade-off between performance and speed:
\begin{itemize}
    \item \textbf{Moment matching} generally outperforms standard Gaussian approximations if an exact solution exists (more so than mean-Hessian matching) and the dimensionality is not too high, but is compromised by an increase in run time with the estimation of the covariance and third moments, requiring importance sampling.
    \item \textbf{Derivative matching} offers an improvement in accuracy to the regular Laplace approximation when there is low data and the posterior is very skewed, at very little additional run time cost.
    \item \textbf{Mean-mode-Hessian matching} provides as or better performance compared to most standard Gaussian approximations when the dimensionality is not too high, at the cost of increased run time
    \item \textbf{Mean-mode-covariance matching} generally outperforms standard Gaussian approximations if an exact solution exists (more so than mean-Hessian matching) and the dimensionality is not too high, at the cost of increased run time with the estimation of the covariance, requiring importance sampling. Exact solutions are more frequent compared to moment matching.
\end{itemize}
The mean-Hessian and mean-covariance matching schemes were additionally shown to be successfully used for post-hoc skewness adjustments for standard Gaussian approximations.
For the case of probit regression, these adjustments were shown to generally result in increased marginal accuracies compared to the base approximation:
\begin{itemize}
    \item \textbf{Mean-mode-Hessian adjustments} tend to give larger increases in marginal accuracies, but also risks decreasing these marginal accuracies when the dimensionality of the data is too large. Better performance was seen when there was low data.
    \item \textbf{Mean-mode-covariance adjustments} tend to give more modest increases in marginal accuracies but generally did not decrease marginal accuracies.
\end{itemize}
Although the skew-normal matching method has its benefits when the posterior is known to be skewed, this is not without its drawbacks.
Using a matching, rather than an optimal variational approach can impact performance for some problems.
This may explain why performance breaks down when the number of dimensions is large.
We are also heavily dependent on good estimates of the moments; without these, the skew-normal approximation can fail very easily.
For high dimensional data where $p\gg n$, skew normal matching is still possible, provided that the appropriate inverses exist to quantities such as the Hessian, although further work needs to be done in verifying the quality of these approximations.

There is clearly a variety of ways the work here could be extended.
Other families of skewed densities could be entertained, the matching methods could be used as warm starts for skew-normal variational approximations \citep{Ormerod2011,SmithEtal2020}, and matching with other posterior statistics is possible.

We hope that the results presented in this paper act as a useful benchmark for the performance of matching-based skewed approximations to posterior distributions.
We believe that the skew-normal matching method fulfills a certain niche in posterior approximations and would like to see similar, if not better, methodologies being developed as approximate Bayesian inference grows and the need for such approximations increase.

\bibliographystyle{plainnat}
\bibliography{bib/main}

\end{document}


\title{Supplementary material for ``Skew-Normal Posterior Approximations''}
\author{Jackson Zhou, Clara Grazian, and John T. Ormerod}
\maketitle

\section{Derivations for the skew-normal matching method}
\label{sec:matching-derivations}

Preliminary results and notation are provided, before each matching scheme is derived.
If the random variable $\vTheta$ has an $\text{SN}_p(\vmu,\mSigma,\vd)$ distribution (as defined in the main text), then it can be shown that the gradient, Hessian, and third-order unmixed derivatives (TUD) of the log density are given by
\begin{align}
    \nabla\log p(\vtheta)&=-\mSigma^{-1}(\vtheta-\vmu)+\zeta_1(\vd^\top(\vtheta - \vmu))\vd,\label{eq:logmsngrad}\\
    \nabla^2\log p(\vtheta)&=-\mSigma^{-1}+\zeta_2(\vd^\top(\vtheta-\vmu))\vd\vd^\top,\ \mbox{and}\label{eq:logmsnhess}\\
    \left(D_{\theta_i}^3\log p(\vtheta)\right)_{i=1,\ldots,p}^\top&=\zeta_3(\vd^\top(\vtheta-\vmu))\vd^{\odot3} \label{eq:logmsntud}
\end{align}
respectively.
Furthermore, recall from the main text that the expectation, variance, and third-order unmixed moments of such a distribution are given by
\begin{align}
    \E(\vTheta)&=\vmu+\sqrt{\frac{2}{\pi}}\vdelta,\label{eq:msnexp}\\
    \var(\vTheta)&=\mSigma - \frac{2}{\pi} \vdelta\vdelta^\top,\ \mbox{and}\label{eq:msnvar}\\
    \text{TUM}(\vTheta)&=\frac{\sqrt{2}(4-\pi)}{\pi^{3/2}}\vdelta^{\odot3}\label{eq:msntum}
\end{align}
respectively, where $\vdelta=\mSigma\vd/\sqrt{1+\vd^\top\mSigma\vd}$ and $\text{TUM}(\vTheta)$ is the vector of third-order unmixed central moments.

\subsection{Moment matching}\label{sec:mmatch}

For $\widetilde{\vmu}$ to be the mean, we use (\ref{eq:msnexp}) and see that we need
\begin{equation}\label{eq:matchmean}
    \widetilde{\vmu}=\vmu+\sqrt{\frac{2}{\pi}}\cdot\frac{\mSigma\vd}{\sqrt{1+\vd^\top\mSigma\vd}}.
\end{equation}
For $\mC$ to be the covariance matrix, we use (\ref{eq:msnvar}) see that we need
\begin{equation}\label{eq:matchcov}
    \mC=\mSigma - \frac{2}{\pi(1+\vd^\top\mSigma\vd)}\mSigma\vd\vd^\top\mSigma.
\end{equation}
Finally, for $\vs$ to be the TUM, we use (\ref{eq:msntum}) and that we need
\begin{equation}\label{eq:matchtum}
    \vs=\frac{\sqrt{2}(4-\pi)}{\pi^{3/2}}\vdelta^{\odot3}.
\end{equation}
Combining (\ref{eq:matchmean}), (\ref{eq:matchcov}), and (\ref{eq:matchtum}), we arrive at the moment matching equations in the main text, which have a straightforward solution.

\subsection{Derivative matching}\label{sec:derivmatch}

For $\vm$ to be the mode, we require that $\nabla\log p(\vm)=\vzero$; by (\ref{eq:logmsngrad}), this is equivalent to
\begin{equation}\label{eq:matchmode}
    \vzero=-\mSigma^{-1}(\vm-\vmu)+\zeta_1(\vd^\top(\vm-\vmu))\vd.
\end{equation}
Next, for $\mJ$ to be the negative Hessian at the mode, we require that $-\nabla^2\log p(\vm)=\mJ$, which by (\ref{eq:logmsnhess}) is equivalent to
\begin{equation}\label{eq:matchhess}
    \mJ=\mSigma^{-1}-\zeta_2(\vd^\top(\vm-\vmu))\vd\vd^\top.
\end{equation}
Finally, for $\vt$ to be the TUD at the mode, we substitute $\vm$ for $\vtheta$ in (\ref{eq:logmsntud}) and set everything to be equal to $\vt$, which gives
\begin{equation}\label{eq:matchtud}
    \vt=\zeta_3(\vd^\top(\vm-\vmu))\vd^{\odot3}.
\end{equation}
Note that $\vd^\top(\vm-\vmu)$ is a recurring quantity in the previous equations.
If we set its value to $\kappa$ and combine (\ref{eq:matchmode}), (\ref{eq:matchhess}), and (\ref{eq:matchtud}), we arrive at the derivative matching equations in the main text, which are
\begin{subequations}\label{eq:derivmatch}
\begin{align}
    \vzero&=-\mSigma^{-1}(\vm-\vmu)+\zeta_1(\kappa)\vd \label{eq:dma}, \\
    \mJ&=\mSigma^{-1}-\zeta_2(\kappa)\vd\vd^\top \label{eq:dmb}, \\
    \vt&=\zeta_3(\kappa)\vd^{\odot3} \label{eq:dmc},\ \mbox{and}\\
    \kappa&=\vd^\top(\vm -\vmu). \label{eq:dmd}
\end{align}
\end{subequations}
The idea is to initially solve for $\kappa$, before recovering the other parameters.
To start, rearrange (\ref{eq:dma}) to get
\begin{equation}\label{eq:vm}
    \vm=\vmu+\zeta_1(\kappa)\mSigma\vd.
\end{equation}
After subtracting $\vmu$ from both sides, left multiplying by $\vd^\top$, and using (\ref{eq:dmd}), we have
\begin{equation}\label{eq:dsd}
    \frac{\kappa}{\zeta_1(\kappa)} = \vd^\top\mSigma\vd.
\end{equation}
Now manipulate (\ref{eq:dmc}) to get
$\vd=\vt^{\odot1/3}/\zeta_3(\kappa)^{1/3}  =\vu/\zeta_3(\kappa)^{1/3}$, 
where we let $\vu=\vt^{\odot1/3}$.
Substituting this into (\ref{eq:dsd}), we get
\begin{equation}\label{eq:usu}
    \frac{\kappa}{\zeta_1(\kappa)}= \frac{1}{\zeta_3(\kappa)^{2/3}}\vu^\top\mSigma\vu
    \implies
    \frac{\kappa\cdot\zeta_3(\kappa)^{2/3}}{\zeta_1(\kappa)}=
    \vu^\top\mSigma\vu.
\end{equation}
If we now rearrange (\ref{eq:dmb}), we have that $\mSigma^{-1}=\mJ+\zeta_2(\kappa)\vd\vd^\top$
after which we can use the Sherman-Morrison identity to arrive at
$$
\mSigma=\mJ^{-1}-\frac{\zeta_2(\kappa)\mJ^{-1}\vd\vd^\top\mJ^{-1}}{1+\zeta_2(\kappa)\vd^\top\mJ^{-1}\vd}.
$$

\noindent 
Left and right multiplying this relation by $\vu$ and substituting in (\ref{eq:usu}), we see that
\begin{align*}
    \frac{\kappa\cdot\zeta_3(\kappa)^{2/3}}{\zeta_1(\kappa)}
    &=\vu^\top\mJ^{-1}\vu-\frac{\zeta_2(\kappa)\vu^\top\mJ^{-1}\vd\vd^\top\mJ^{-1}\vu}{1+\zeta_2(\kappa)\vd^\top\mJ^{-1}\vd} \\
    \implies\frac{\kappa\cdot\zeta_3(\kappa)^{2/3}}{\zeta_1(\kappa)}
    &=\vu^\top\mJ^{-1}\vu-\frac{\zeta_2(\kappa)}{\zeta_3(\kappa)^{2/3}}\cdot\frac{\vu^\top\mJ^{-1}\vu\vu^\top\mJ^{-1}\vu}{1+\frac{\zeta_2(\kappa)}{\zeta_3(\kappa)^{2/3}}\vu^\top\mJ^{-1}\vu}.
\end{align*}
Further substituting in $R=\vu^\top\mJ^{-1}\vu$, the above equation becomes
$$
    \frac{\kappa\cdot\zeta_3(\kappa)^{2/3}}{\zeta_1(\kappa)}=R-\frac{\zeta_2(\kappa)}{\zeta_3(\kappa)^{2/3}}\cdot\frac{R^2}{1+\frac{\zeta_2(\kappa)}{\zeta_3(\kappa)^{2/3}}R} \\[6pt]
    \implies\frac{\kappa\cdot\zeta_3(\kappa)^{2/3}}{\zeta_1(\kappa)}=R-\frac{\zeta_2(\kappa)R^2}{\zeta_3(\kappa)^{2/3}+\zeta_2(\kappa)R}.
$$
This final equation is univariate in $\kappa$ and so can be solved numerically.
Once $\kappa$ is known, (\ref{eq:dmc}) can be used to find $\vd$, after which (\ref{eq:dmb}) can be used to find $\mSigma$.
Finally, (\ref{eq:vm}) can be used to find $\vmu$.

\subsection{Mean-mode-Hessian matching}\label{sec:mhmatch}

For $\vm$ to be the mode, we use our working from Section \ref{sec:derivmatch} and see that we need (\ref{eq:matchmode}) to be satisfied.
For $\mJ$ to be the negative Hessian at the mode, we use our working from Section \ref{sec:derivmatch} and see that we need (\ref{eq:matchhess}) to be satisfied.
Finally, for $\widetilde{\vmu}$ to be the mean, we use our working from Section \ref{sec:mmatch} and see that we need (\ref{eq:matchmean}) to be satisfied.
Combining (\ref{eq:matchmode}), (\ref{eq:matchhess}), and (\ref{eq:matchmean}), we arrive at the mean-mode-Hessian matching equations in the main text, which are
\begin{subequations}\label{eq:mHmatch}
\begin{align}
    \vzero&=-\mSigma^{-1}(\vm-\vmu)+\zeta_1(\kappa)\vd, \label{eq:mhma} \\
    \mJ&=\mSigma^{-1}-\zeta_2(\kappa)\vd\vd^\top, \label{eq:mhmb} \\
    \widetilde{\vmu}&=\vmu+\sqrt{\frac{2}{\pi}}\cdot\frac{\mSigma\vd}{\sqrt{1+\vd^\top\mSigma\vd}},\ \mbox{and} \label{eq:mhmc} \\
    \kappa&=\vd^\top(\vm -\vmu). \label{eq:mhmd}
\end{align}
\end{subequations}
Similar to before, the idea is to first solve for $\kappa$, before recovering the other parameters.
As with Section \ref{sec:derivmatch}, (\ref{eq:mhma}), and (\ref{eq:mhmd}) can be used to deduce (\ref{eq:vm}) and (\ref{eq:dsd}).
If we subtract (\ref{eq:mhmc}) by (\ref{eq:vm}) and substitute (\ref{eq:dsd}), we get that
\begin{equation}\label{eq:vDelta}
    \vDelta=\widetilde{\vmu}-\vm=\lambda(\kappa)\mSigma\vd,\quad\mbox{with}\quad\lambda(\kappa) = \sqrt{\frac{2}{\pi}}\cdot\frac{1}{\sqrt{1 + \frac{\kappa}{\zeta_1(\kappa)}}} - \zeta_1(\kappa).
\end{equation}
Substituting into (\ref{eq:mhmb}), we have
$\mJ=\mSigma^{-1} - (\zeta_2(\kappa)/\lambda(\kappa)^2)  \mSigma^{-1}\vDelta\vDelta^\top \mSigma^{-1}$.
Applying the Sherman-Morrison identity to this relation, we get that
\begin{equation}\label{eq:smw1}
    \mJ^{-1} = \mSigma + \frac{\vDelta\vDelta^\top}{\lambda(\kappa)^2/ \zeta_2(\kappa)  - \vDelta^\top\mSigma^{-1}\vDelta }.
\end{equation}
Now, if we rearrange (\ref{eq:mhmb}), we see that $\mSigma^{-1}  = \mJ + \zeta_2(\kappa) \vd\vd^\top$.
Left and right multiplying this by $\vDelta$, we have
\begin{equation*}
\begin{array}{rl}
    \vDelta^\top\mSigma^{-1}\vDelta&=\vDelta^\top\mJ\vDelta + \zeta_2(\kappa) (\vd^\top\vDelta)^2 \\
    &=\vDelta^\top\mJ\vDelta + \zeta_2(\kappa) \lambda(\kappa)^2 (\vd^\top\mSigma\vd)^2 \\
    &=\vDelta^\top\mJ\vDelta + \zeta_2(\kappa) \lambda(\kappa)^2 \left( \frac{\kappa}{\zeta_1(\kappa)} \right)^2.
\end{array} 
\end{equation*}
Substituting this into (\ref{eq:smw1}) gives us
\begin{equation*}
    \mJ^{-1} = \mSigma + \frac{\vDelta\vDelta^\top}{\lambda(\kappa)^2/ \zeta_2(\kappa)  - \vDelta^\top\mJ\vDelta - \zeta_2(\kappa) \lambda(\kappa)^2 \left( \frac{\kappa}{\zeta_1(\kappa)} \right)^2 } = \mSigma + \alpha(\kappa) \vDelta\vDelta^\top,
\end{equation*}
where
\begin{equation*}
    \alpha(\kappa) = \left(\lambda(\kappa)^2/ \zeta_2(\kappa)  - \vDelta^\top\mJ\vDelta - \zeta_2(\kappa) \lambda(\kappa)^2 \left( \frac{\kappa}{\zeta_1(\kappa)} \right)^2\right)^{-1}.
\end{equation*}

\noindent 
Hence, we get that
\begin{equation}\label{eq:Sigma}
    \mSigma = \mJ^{-1} - \alpha(\kappa)\vDelta\vDelta^\top\implies\mSigma^{-1} = \mJ + \frac{\mJ\vDelta\vDelta^\top\mJ}{1/\alpha(\kappa) - \vDelta^\top\mJ\vDelta},
\end{equation}
where the Sherman-Morrison identity was used for the implication.
Substituting this into (\ref{eq:vDelta}), we have
$$
\ds \vd=\frac{\mSigma^{-1}\vDelta}{\lambda(\kappa)}
=\frac{1}{\lambda(\kappa)}\left[\mJ+\frac{\mJ\vDelta\vDelta^\top\mJ}{1/\alpha(\kappa)-\vDelta^\top\mJ\vDelta}\right]\vDelta
=\frac{1}{\lambda(\kappa)}\left[1+\frac{\vDelta^\top\mJ\vDelta}{1/\alpha(\kappa)-\vDelta^\top\mJ\vDelta}\right]\mJ\vDelta.
$$

\noindent 
Substituting again into (\ref{eq:Sigma}), we are left with
\begin{equation*}
    \vd^\top\mSigma\vd = \frac{1}{\lambda(\kappa)^2}\left[1 +  \frac{ \vDelta^\top\mJ\vDelta}{1/\alpha(\kappa) - \vDelta^\top\mJ\vDelta} \right]^2 \vDelta^\top\mJ \left[ \mJ^{-1} - \alpha(\kappa)\vDelta\vDelta^\top \right] \mJ\vDelta=\frac{\kappa}{\zeta_1(\kappa)},
\end{equation*}
where (\ref{eq:dsd}) was used for the second equality.
By letting $Q=\vDelta^\top\mJ\vDelta$ and rearranging the latter part of the previous equation, we see that
\begin{align*}
    \left[1+\frac{Q}{1/\alpha(\kappa)-Q}\right]^2\vDelta^\top\mJ \left[ \mJ^{-1} - \alpha(\kappa)\vDelta\vDelta^\top \right] \mJ\vDelta&=\frac{\kappa\lambda(\kappa)^2}{\zeta_1(\kappa)} \\
    \implies\left[\frac{1}{1-\alpha(\kappa)Q}\right]^2\vDelta^\top\left[ \mJ - \alpha(\kappa)\mJ\vDelta\vDelta^\top\mJ \right]\vDelta&=\frac{\kappa\lambda(\kappa)^2}{\zeta_1(\kappa)} \\
    \implies\left[\frac{1}{1-\alpha(\kappa)Q}\right]^2\left[Q-\alpha(\kappa)Q^2\right]&=\frac{\kappa\lambda(\kappa)^2}{\zeta_1(\kappa)} \\
    \implies\frac{Q}{1-\alpha(\kappa)Q}&=\frac{\kappa\lambda(\kappa)^2}{\zeta_1(\kappa)} \\
    \implies\frac{1}{Q}-\alpha(\kappa)-\frac{\zeta_1(\kappa)}{\kappa\lambda(\kappa)^2}&=0.
\end{align*}
As with Section \ref{sec:derivmatch}, this is a univariate equation in $\kappa$ and so can be solved numerically.
Once $\kappa$ is known, we can use (\ref{eq:Sigma}) to solve for $\mSigma$, after which we can use (\ref{eq:vDelta}) to solve for $\vd$.
Finally, we can substitute everything into (\ref{eq:mhmc}) to solve for $\vmu$.

\subsection{Mean-mode-covariance matching}\label{sec:mcmatch}

For $\vm$ to be the mode, we use our working from Section \ref{sec:derivmatch} and see that we need (\ref{eq:matchmode}) to be satisfied.
For $\widetilde{\vmu}$ to be the mean, we use our working from Section \ref{sec:mmatch} and see that we need (\ref{eq:matchmean}) to be satisfied.
Finally, for $\mC$ to be the mode, we use our working from Section \ref{sec:mmatch} and see that we need (\ref{eq:matchcov}) to be satisfied.
Combining (\ref{eq:matchmode}), (\ref{eq:matchmean}), and (\ref{eq:matchcov}), we arrive at the mean-mode-covariance matching equations in the main text, which are
\begin{subequations}\label{eq:mcmatch}
\begin{align}
    \vzero&=-\mSigma^{-1}(\vm-\vmu)+\zeta_1(\kappa)\vd, \label{eq:mcma} \\
    \widetilde{\vmu}&=\vmu+\sqrt{\frac{2}{\pi}}\cdot\frac{\mSigma\vd}{\sqrt{1+\vd^\top\mSigma\vd}}, \label{eq:mcmb} \\
    \mC&=\mSigma - \frac{2}{\pi(1+\vd^\top\mSigma\vd)}\mSigma\vd\vd^\top\mSigma,\ \mbox{and} \label{eq:mcmc} \\
    \kappa&=\vd^\top(\vm -\vmu). \label{eq:mcmd}
\end{align}
\end{subequations}
Similar to before, the idea is to first solve for $\kappa$, before recovering the other parameters.
As with Section \ref{sec:mhmatch}, (\ref{eq:mcma}), (\ref{eq:mcmb}), and (\ref{eq:mcmd}) can be used to deduce (\ref{eq:dsd}) and (\ref{eq:vDelta}).
Substituting both relations into (\ref{eq:mcmc}), we are left with
\begin{equation}
    \mC=\mSigma-\frac{2}{\pi\left(1+\frac{\kappa}{\zeta_1(\kappa)}\right)}\cdot\frac{1}{\lambda(\kappa)^2}\vDelta\vDelta^\top
    \quad \implies \quad \mSigma=\mC+\beta(\kappa)\vDelta\vDelta^\top, \label{eq:mC}
\end{equation}
where
\begin{equation}\label{eq:betafun}
    \beta(\kappa)=\frac{2}{\pi\left(1+\frac{\kappa}{\zeta_1(\kappa)}\right)}\cdot\frac{1}{\lambda(\kappa)^2}.
\end{equation}
Note that $\lambda(\kappa)$ refers to the same function in (\ref{eq:vDelta}).
By applying the Sherman-Morrison identity, this becomes
$\mSigma^{-1}=\mC^{-1}-\mC^{-1}\vDelta\vDelta^\top\mC^{-1}/[\beta(\kappa)^{-1}+\vDelta^\top\mC^{-1}\vDelta]$.
Substituting this result back into (\ref{eq:vDelta}), we see that
$$
\ds \vd 
=\frac{1}{\lambda(\kappa)}\mSigma^{-1}\vDelta 
=\frac{1}{\lambda(\kappa)}\left[1-\frac{\vDelta^\top\mC^{-1}\vDelta}{\frac{1}{\beta(\kappa)}+\vDelta^\top\mC^{-1}\vDelta}\right]\mC^{-1}\vDelta.
$$

\noindent 
Combining this with (\ref{eq:mC}), we now get
\begin{equation*}
    \vd^\top\mSigma\vd=\frac{1}{\lambda(\kappa)^2}\left[1-\frac{\vDelta^\top\mC^{-1}\vDelta}{\frac{1}{\beta(\kappa)}+\vDelta^\top\mC^{-1}\vDelta}\right]^2\vDelta^\top\mC^{-1}\left[\mC+\beta(\kappa)\vDelta\vDelta^\top\right]\mC^{-1}\vDelta=\frac{\kappa}{\zeta_1(\kappa)},
\end{equation*}
where (\ref{eq:dsd}) was used for the second equality.
By letting $G=\vDelta^\top\mC^{-1}\vDelta$ and rearranging the latter part of the previous equation, we see that
\begin{align*}
    \left[1-\frac{G}{\frac{1}{\beta(\kappa)}+G}\right]^2\left[G+\beta(\kappa)G^2\right]&=\frac{\kappa\lambda(\kappa)^2}{\zeta_1(\kappa)} \\
    \implies\left[\frac{1}{1+\beta(\kappa)G}\right]^2G\left[1+\beta(\kappa)G\right]&=\frac{\kappa\lambda(\kappa)^2}{\zeta_1(\kappa)} \\
    \implies\frac{G}{1+\beta(\kappa)G}&=\frac{\kappa\lambda(\kappa)^2}{\zeta_1(\kappa)} \\
    \implies\frac{1}{G}+\beta(\kappa)-\frac{\zeta_1(\kappa)}{\kappa\lambda(\kappa)^2}&=0.
\end{align*}
As with the previous sections, this is a univariate equation in $\kappa$ and so can be solved numerically.
Once $\kappa$ is known, we can use (\ref{eq:mC}) to solve for $\mSigma$, after which (\ref{eq:vDelta}) can be used to solve for $\vd$.
Finally, we can substitute everything into (\ref{eq:mcmb}) to solve for $\vmu$.

\subsection{On solving the kappa equations}

In the derivative matching, mean-mode-Hessian and mean-mode-covariance matching schemes, it was seen empirically that the graph of the final equation in $\kappa$ had a hyperbolic structure, and when a root of the graph did exist, it was always present on the positive half of the real line.
This requires further investigation, but for now we only considered strictly positive $\kappa$ values when searching for a solution in our implementation.

Additionally, for the mean-mode-covariance matching scheme in particular, it can be shown that the horizontal asymptote for the positive half of the graph in $\kappa$, minus the $1/G$ term depending on the data, is at $1-\pi/2$.
This implies that when $G\geq2/(\pi-2)$, there is no solution to (\ref{eq:mcmatch}).
To see this fact about the horizontal asymptote, we consider the limit of $\beta(\kappa)-\zeta_1(\kappa)/(\kappa\lambda(\kappa)^2)$ as $\kappa\to\infty$.
The second term can expanded out using the definition of $\lambda(\kappa)$, and has the form
\begin{equation*}
    \frac{\zeta_1(\kappa)}{\kappa\lambda(\kappa)^2}=\frac{\zeta_1(\kappa)/\kappa + 1}{\left(\sqrt{\frac{2}{\pi}}-\sqrt{\zeta_1(\kappa)^2+\kappa\zeta_1(\kappa)}\right)^2}.
\end{equation*}
As $\kappa\to\infty$, the numerator is seen to have a limit of 1, while the denominator is seen to have a limit of $2/\pi$, and so the second term has a limit of $\pi/2$ as $\kappa\to\infty$.
Applying this result to the definition of $\beta(\kappa)$, it also becomes clear that $\beta(\kappa)\to1$ as $\kappa\to\infty$.
Combining both parts gives us our original result.

\section{Estimation of key statistics}
\label{sec:statistic-estimation}

In the skew-normal matching method, key statistics of the observed posterior $p(\vtheta|\sD)$ need to be estimated before they can be matched to the multivariate skew-normal density.
The following subsections describe how these statistics were estimated.

\subsection{Estimation of derivatives at the mode}

Denote in this and the following sections the observed log joint likelihood by $f(\vtheta)$, with the corresponding gradient and Hessian functions given by $\vg(\vtheta)$ and $\mH(\vtheta)$ respectively.
Perhaps the simplest approach in obtaining the posterior mode is the Newton-Raphson method, described in detail by \cite{ypma1995historical}.
Newton-Raphson updates of the form
\begin{equation*}
    \vm_{n+1} \leftarrow \vm_n - \mH(\vm_n)^{-1}\vg(\vm_n)
\end{equation*}
allow for the mode of the observed posterior to be determined relatively quickly, after which other higher-order derivatives at the mode can be evaluated analytically.
Convergence is typically defined to be the point when $\norm{\vg(\vm_{n})}<t$, for some positive threshold $t$.
As the Newton-Raphson method only finds regions with zero gradient, care needs to be taken to ensure that the global maximum is reached, rather than a local maximum.
This will always be the case if the posterior is unimodal.
Additionally, note that in practice, Newton-Raphson updates which are too large can lead to non-convergence in the algorithm, where updates do not necessarily increase the objective function.
To remedy this, updates was modified to have the dampened form:
$\vm_{n+1} \leftarrow \vm_n - \lambda\cdot\mH(\vm_n)^{-1}\vg(\vm_n)$.
In each iteration, $\lambda\in(0,1]$ is chosen via a simplified one-dimensional line search so as to ensure an increase in the objective function.

\subsection{Estimation of moments}

\subsubsection{Importance sampling}

All moments are able to be estimated via importance sampling if there is no other alternative.
In keeping with the spirit of the skew-normal matching method, we chose the proposal density $g$ to be an appropriate $t$-distribution-based approximation of the posterior, so as to ensure proper coverage of the posterior tails.
This approximation has the form of a location-shifted multivariate $t$-distribution, as described by \cite{genz_bretz_2009}.
The shift and scale parameters are set to be $\vm$ and $\mJ$ respectively, which will have been estimated from the previous section.
The degrees of freedom for this $t$-distribution were chosen so as to encourage finite variances during the importance sampling stage.
Finally, stabilizing adjustments to the important ratios \citep{vehtari2021} were made to alleviate noisy estimates.

\subsubsection{Alternate estimates for the mean}

If the statistic of interest is only the first moment, then there readily exist alternate estimation approaches which may be faster than importance sampling.
For example, one may approximate the marginals of the observed posterior using Jensen's inequality via
\begin{equation}\label{eq:mest}
    p(\theta_j|\sD)\appropto\exp\left[\mathbb{E}_{\vtheta_{-j}\sim q}\left[f(\vtheta)-\log q(\vtheta_{-j})\right]\right]\eqqcolon\hat{p}_\text{Jens.}(\theta_j|q),
\end{equation}
where $q(\vtheta_{-j})$ is some suitable approximation of the conditional posterior $p(\vtheta_{-j}|\theta_j,\sD)$.
In our simulations, we chose $q$ to be the corresponding conditional of the Laplace approximation of the observed posterior.
These marginal estimates can then be normalized via univariate quadrature, after which the marginal means of the posterior can be estimated using the same method.
Finally, these can be combined to form an overall estimate of the mean.
Call this estimate the Jensen-based mean estimate.

Another similar approach would be to instead use the improved Laplace method in obtaining approximations of the posterior marginal.
The original derivations are outlined by \cite{tierney_kadane_1986}; these approximations will be of the form
\begin{equation}
    p(\theta_j|\sD)\appropto|\mSigma_{\theta_j}|^{1/2}\cdot p(\vmu_{\theta_j},\theta_j,\sD)\eqqcolon\hat{p}_\text{IL}(\theta_j|q),
\end{equation}
where $\vmu_{\theta_j}$ and $\mSigma_{\theta_j}$ are the mean and covariance respectively of the Laplace approximation of the joint density $p(\vtheta_{-j},\theta_j,\sD)$, where $\theta_j$ is fixed.
As before, these marginal estimates can then be normalized via univariate quadrature, after which the marginal means of the posterior can be estimated using the same method.
Finally, these can be combined to form an overall estimate of the mean.
Call this the improved Laplace mean estimate.

\section{Probit regression supplement}
\label{sec:probit-supplement}

\subsection{Derivations for the Jensen variant}

We have that $Y_i\sim\text{Bernoulli}(p_i)$ are independent for $i=1,\ldots,n$, where
$p_i=\Phi(\vx_i^T\vtheta)$,
and $\vx_i$ and $\vtheta$ are $\mathbb{R}^p$.
The likelihood function for this probit regression model with observed values $y_1,\ldots,y_n$ and corresponding predictors $\vx_1,\ldots,\vx_n$ is given by
$$
p(\sD|\vtheta)=\prod_{i=1}^n\left[\Phi\left(\vx_i^T\vtheta\right)^{y_i}\left(1-\Phi\left(\vx_i^T\vtheta\right)\right)^{1-y_i}\right].
$$
An alternate parameterization which simplifies calculations is given by
$$
p(\sD|\vtheta)=\prod_{i=1}^n\left[\Phi\left((2y_i-1)\vx_i^T\vtheta\right)\right] =\prod_{i=1}^n\left[\Phi\left(\vz_i^T\vtheta\right)\right],
$$
where $\vz_i=(2y_i-1)\vx_i$ is known; this will be used throughout this section.
The corresponding log-likelihood is $\log p(\sD|\vtheta) = \vone_n^T\zeta_0\left(\mZ\vtheta\right)$
where $\mZ=[\vz_1^T,\ldots,\vz_n^T]^T$.
The log joint likelihood in this case can therefore be written as
\begin{equation}\label{eq:probitf}
    f(\vtheta)=\vone_n^T\zeta_0\left(\mZ\vtheta\right)-\frac{\norm{\vtheta}^2}{2\sigma_\vtheta^2}+\text{constants in $\vtheta$},
\end{equation}
with corresponding gradient and Hessian functions given by
\begin{align}
    \vg(\vtheta)&=\mZ^T\zeta_1(\mZ\vtheta)-\frac{\vtheta}{\sigma_\vtheta^2}\quad\mbox{and}\label{eq:probitg} \\
    \mH(\vtheta)&=\mZ^T\diag(\zeta_2(\mZ\vtheta))\mZ-\frac{1}{\sigma_\vtheta^2}\mI_p.\label{eq:probitH}
\end{align}
In order to implement the Jensen variant of the mean-mode-Hessian matching scheme, we need to evaluate $\hat{p}_\text{Jens.}(\theta_j|q)$ as given in (\ref{eq:mest}), with $q$ being set to the $\vtheta_{-j}$ conditional distribution (keeping $\theta_j$ fixed) of the Laplace approximation of the posterior.
Call this conditional distribution $\sN_{p-1}(\vmu_{\theta_j},\mSigma_{\theta_j})$ (noting that the conditional is also Gaussian).
We have that
\begin{equation*}
\begin{array}{rl} 
    \hat{p}_\text{Jens.}(\theta_j|q)&=\exp\left\{\mathbb{E}_{\vtheta_{-j}\sim q}\left[f(\vtheta)-\log q(\vtheta_{-j})\right]\right\} \\
    &\propto\exp\left\{\mathbb{E}_{\vtheta_{-j}\sim q}\left[\vone_n^T\zeta_0\left(\mZ\vtheta\right)-\frac{\norm{\vtheta}^2}{2\sigma_\vtheta^2}-\log q(\vtheta_{-j})\right]\right\} \\
    &\propto\exp\left\{\mathbb{E}_{\vtheta_{-j}\sim q}\left[\vone_n^T\zeta_0(\mZ\vtheta)\right]-\mathbb{E}_{\vtheta_{-j}\sim q}\left[\frac{\norm{\vtheta}^2}{2\sigma_\vtheta^2}\right]+\frac{1}{2}\log\left|\mSigma_{\theta_j}\right|\right\}.
\end{array} 
\end{equation*}
We will compute each of these expectations individually.
To start, it is clear that
$$
\ds \mathbb{E}_{\vtheta_{-j}\sim q}\left[\frac{\norm{\vtheta}^2}{2\sigma_\vtheta^2}\right]
=\frac{1}{2\sigma_\vtheta^2}\cdot\mathbb{E}_{\vtheta_{-j}\sim q}\left[\theta_1^2+\ldots+\theta_n^2\right]  
=\frac{1}{2\sigma_\vtheta^2}\left[\theta_j^2+\vone_p^\top\left(\vmu_{\theta_j}^{\odot2}+\dg(\mSigma_{\theta_j})\right)\right].
$$
For the first term, a method involving a Taylor expansion and the closure of the Gaussian distribution under affine transformations can be used to find the expectation; we have
\begin{align*}
    \mathbb{E}_{\vtheta_{-j}\sim q}\left[\vone_n^T\zeta_0(\mZ\vtheta)\right]\approx\vone_n^T\left[\zeta_0(\theta_j\mZ_{,j}+\widetilde{\vmu}_{j,\theta_j})+\frac{1}{2}\zeta_2(\theta_j\mZ_{,j}+\widetilde{\vmu}_{j,\theta_j})\odot\widetilde{\vsigma}_{j,\theta_j}^2\right],
\end{align*}
where
$\widetilde{\vmu}_{j,\theta_j}=\mZ_{,-j}\vmu_{\theta_j}$
    and
$\widetilde{\vsigma}_{j,\theta_j}^2 =\dg\left(\mZ_{,-j}\mSigma_{\theta_j}\mZ_{,-j}^T\right)$.
The final form for the Jensen-based approximation is therefore given by
\begin{align*}
    \hat{p}_\text{Jens.}(\theta_j|q)\propto\exp\bigg\{&\vone_n^T\left[\zeta_0(\theta_j\mZ_{,j}+\widetilde{\vmu}_{j,\theta_j})+\frac{1}{2}\zeta_2(\theta_j\mZ_{,j}+\widetilde{\vmu}_{j,\theta_j})\odot\widetilde{\vsigma}_{j,\theta_j}^2\right] \\
    &-\frac{1}{2\sigma_\vtheta^2}\left[\theta_j^2+\vone_p^T\left(\vmu_{\theta_j}^{\odot2}+\dg(\mSigma_{\theta_j})\right)\right]+\frac{1}{2}\log\left|\mSigma_{\theta_j}\right|\bigg\}.
\end{align*}

\subsection{Supplementary results}

\begin{figure}
    \begin{center}
        \includegraphics[scale=0.58]{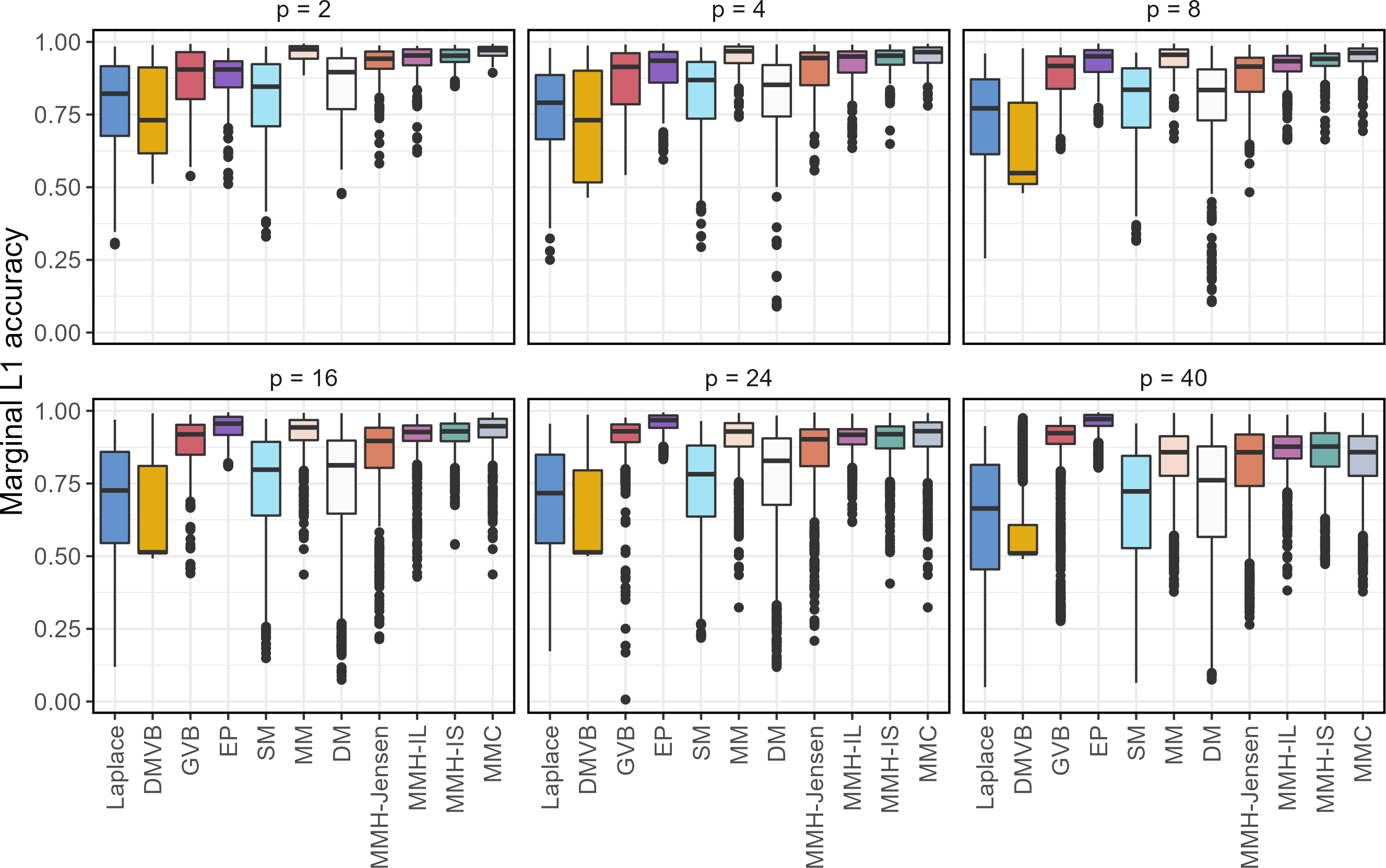}
    \end{center}
    \caption{Performance of approximation methods on simulated independent data across different dimensions for probit regression, with $n=2p$ (box plot).}
    \label{STA-sim-PR-2A-box}
\end{figure}

\begin{figure}
    \begin{center}
        \includegraphics[scale=0.58]{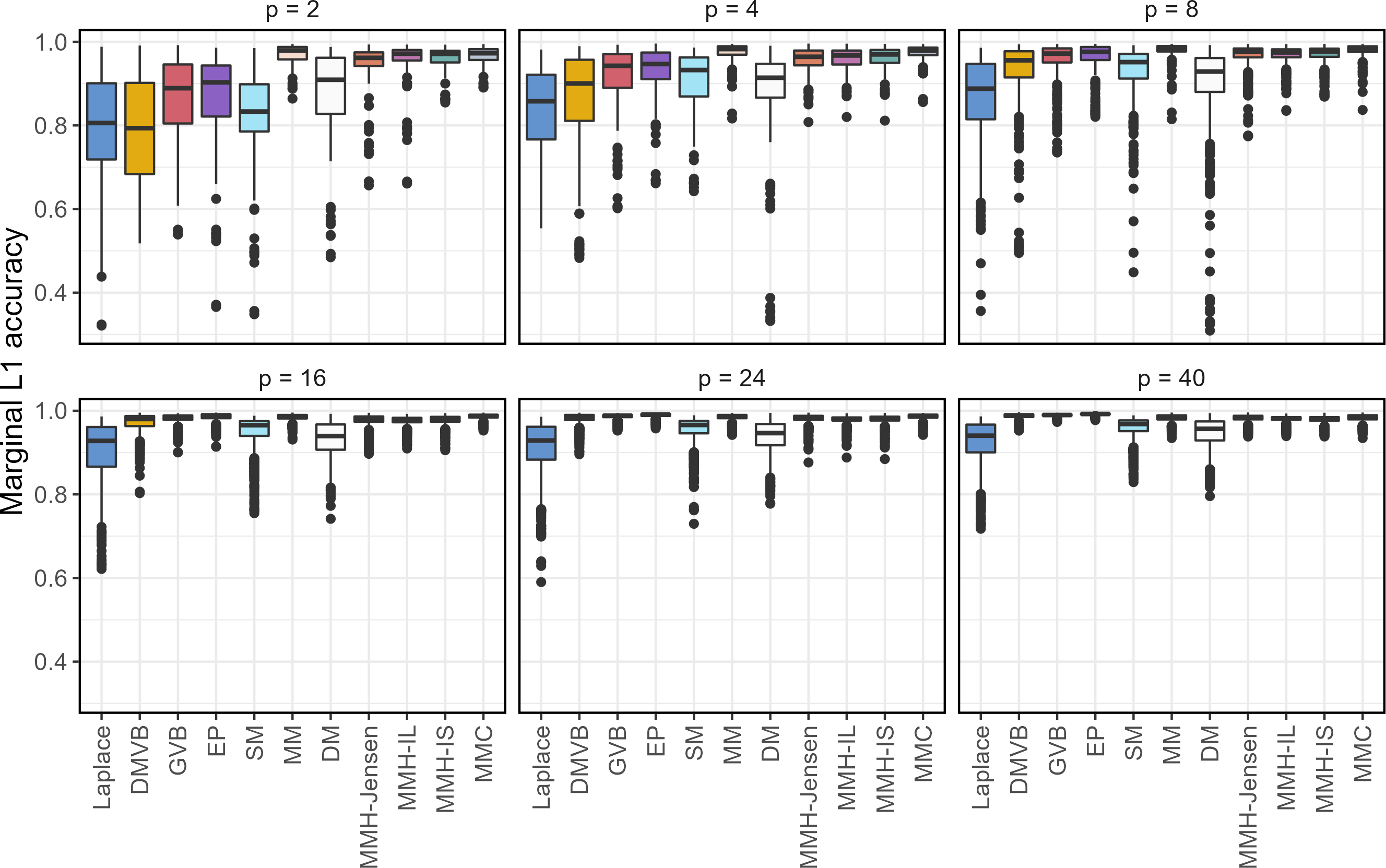}
    \end{center}
    \caption{Performance of approximation methods on simulated independent data across different dimensions for probit regression, with $n=4p$ (box plot).}
    \label{STA-sim-PR-4A-box}
\end{figure}

\begin{figure}
    \begin{center}
        \includegraphics[scale=0.58]{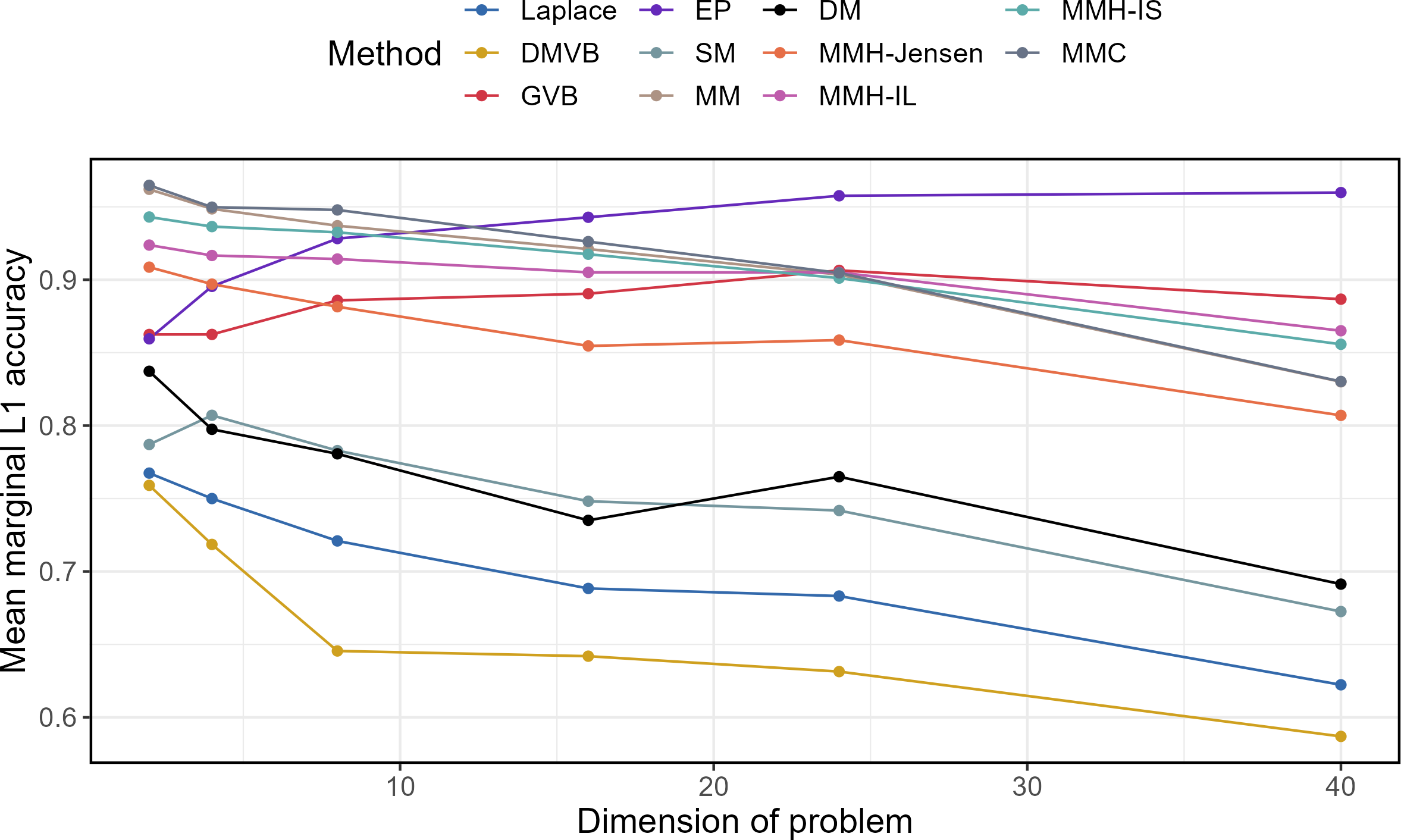}
    \end{center}
    \caption{Performance of approximation methods on simulated independent data across different dimensions for probit regression, with $n=2p$ (line graph).}
    \label{STA-sim-PR-2A-line}
\end{figure}

\begin{figure}
    \begin{center}
        \includegraphics[scale=0.58]{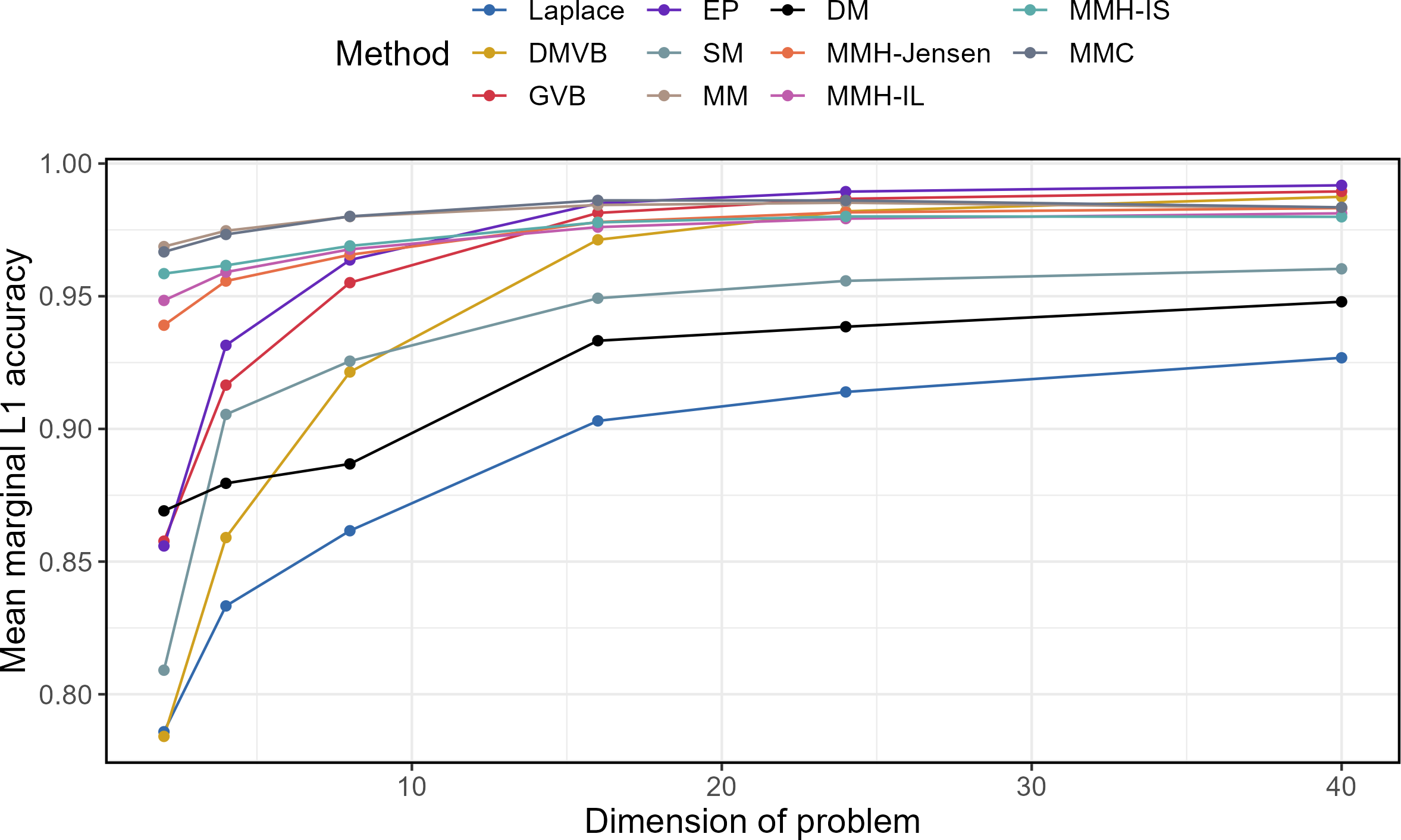}
    \end{center}
    \caption{Performance of approximation methods on simulated independent data across different dimensions for probit regression, with $n=4p$ (line graph).}
    \label{STA-sim-PR-4A-line}
\end{figure}

\begin{figure}
    \begin{center}
        \includegraphics[scale=0.62]{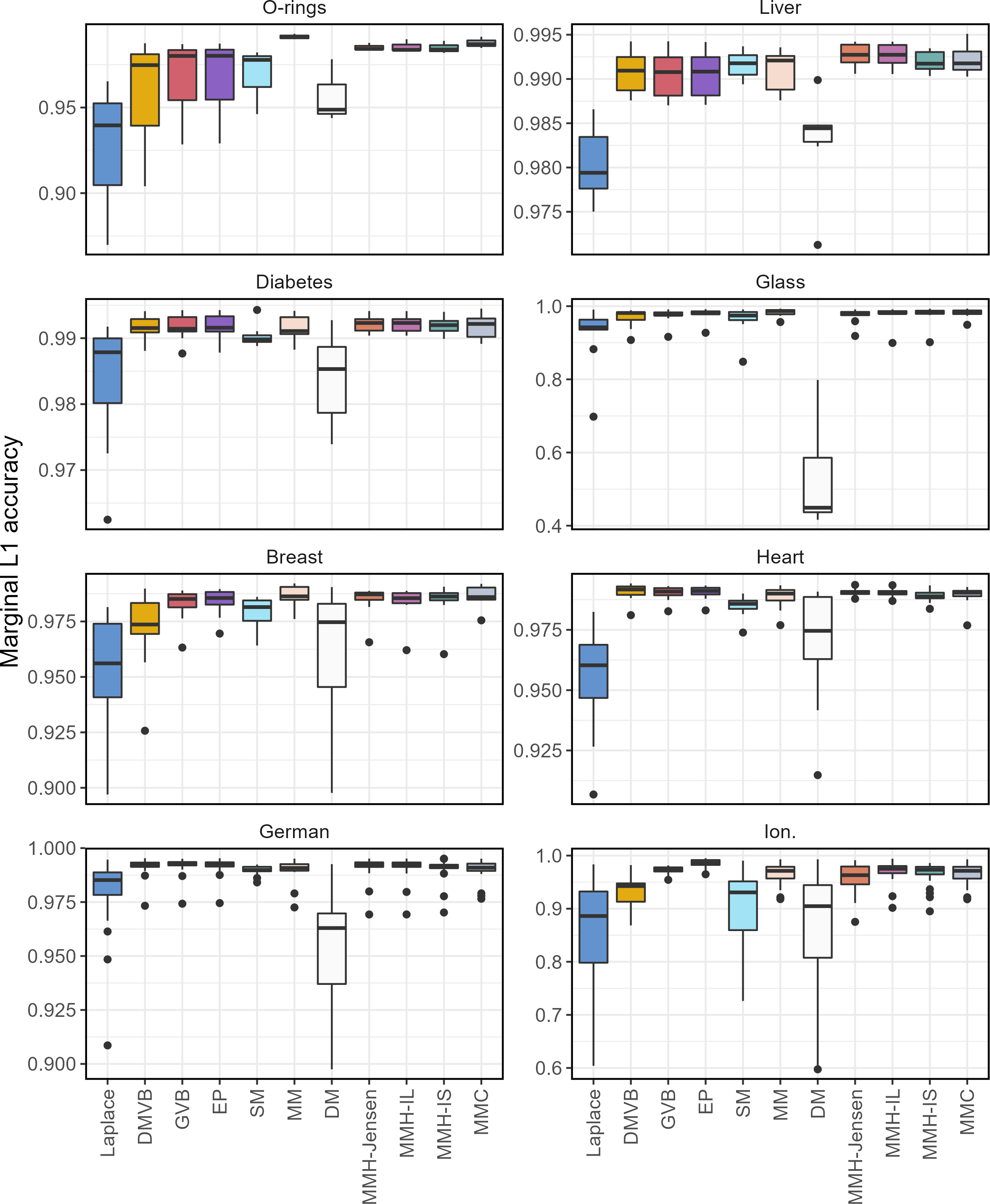}
    \end{center}
    \caption{Performance of approximation methods on benchmark datasets under probit regression.}
    \label{STA-bench-PR-box}
\end{figure}

\begin{figure}
    \begin{center}
        \includegraphics[scale=0.62]{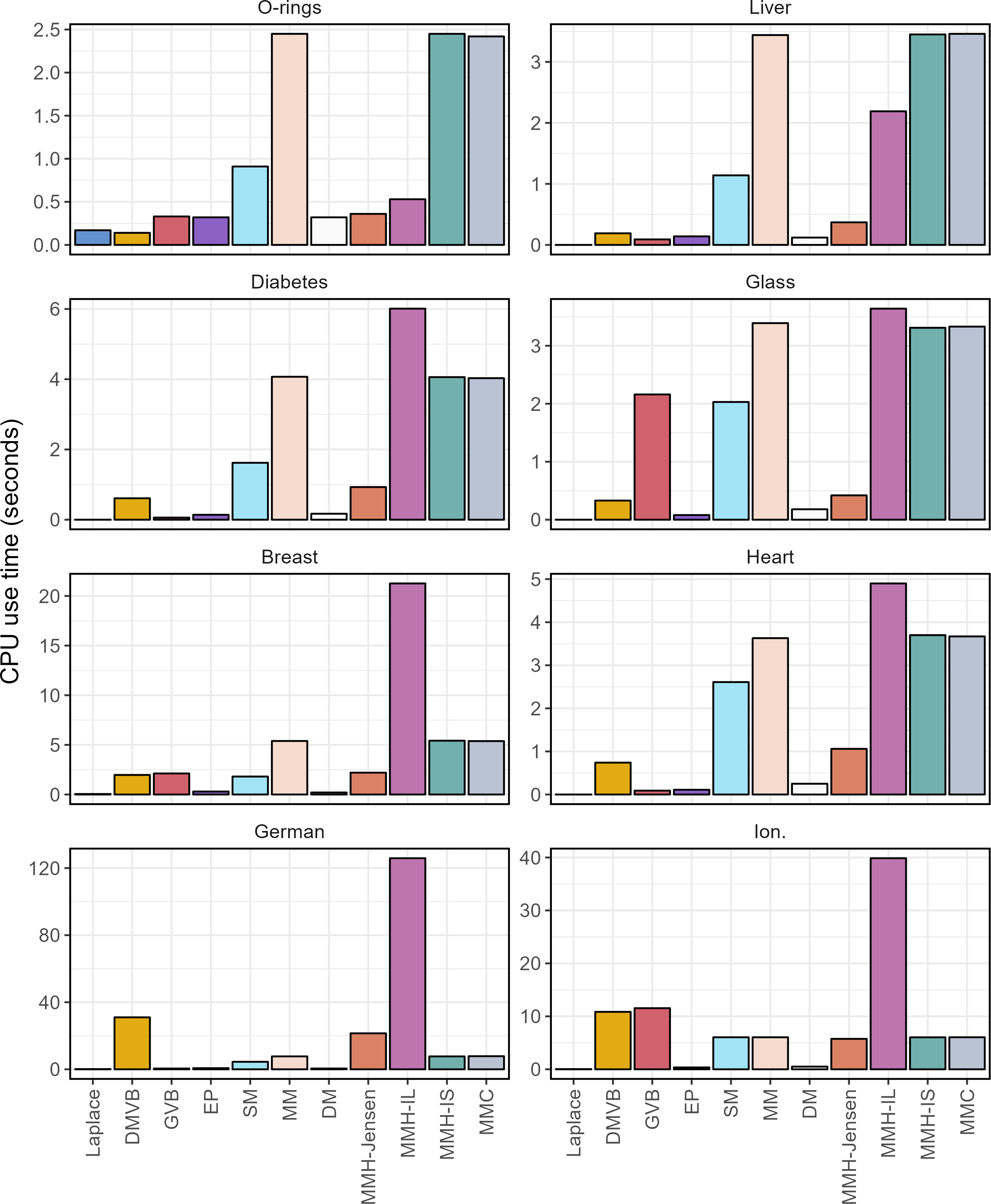}
    \end{center}
    \caption{Time usage of approximation methods on benchmark datasets under probit regression.}
    \label{STA-bench-PR-time}
\end{figure}

\begin{figure}
    \begin{center}
        \includegraphics[scale=0.58]{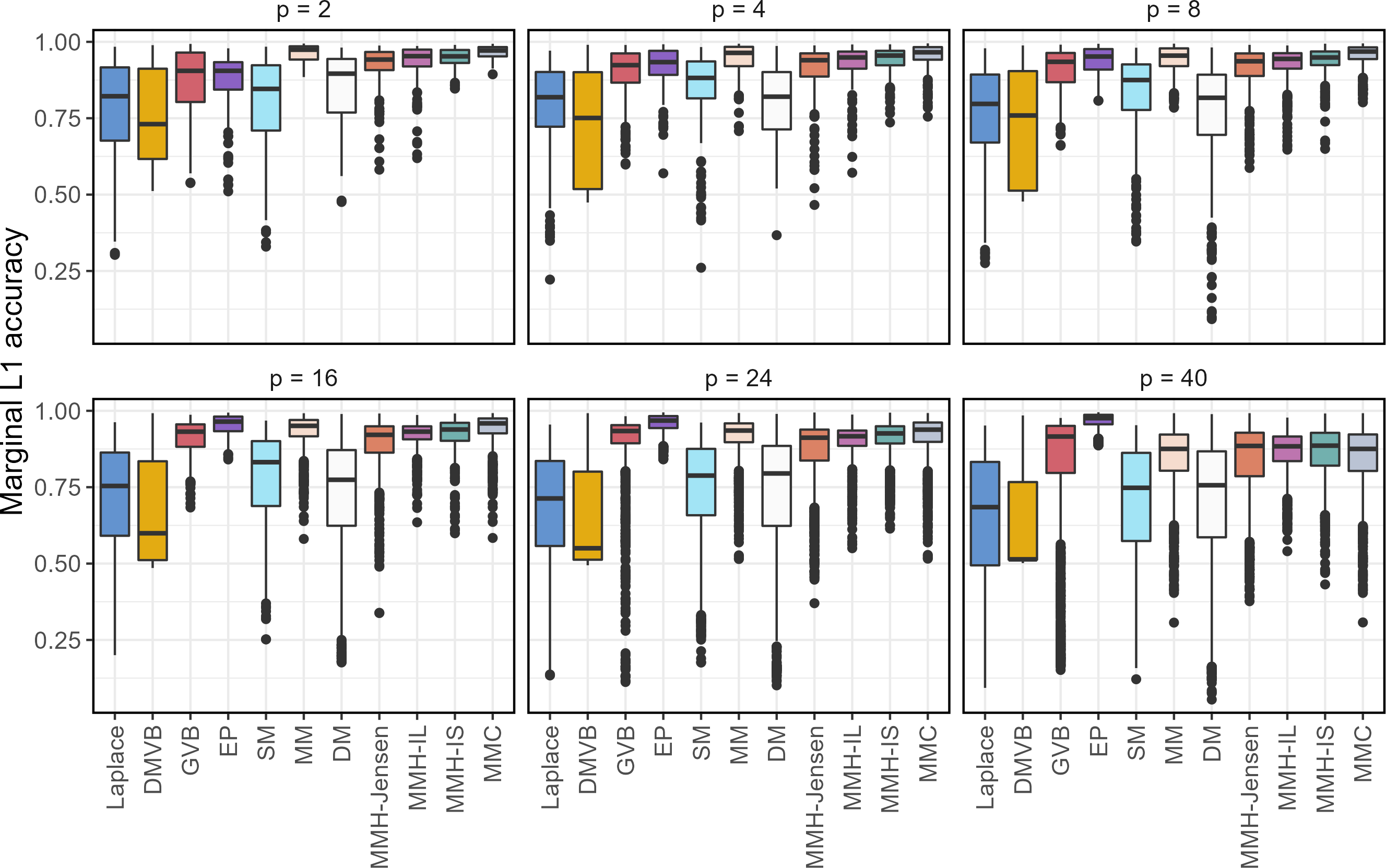}
    \end{center}
    \caption{Performance of approximation methods on simulated AR1 data across different dimensions for probit regression, with $n=2p$ (box plot).}
    \label{STA-sim-PR-2B-box}
\end{figure}

\begin{figure}
    \begin{center}
        \includegraphics[scale=0.58]{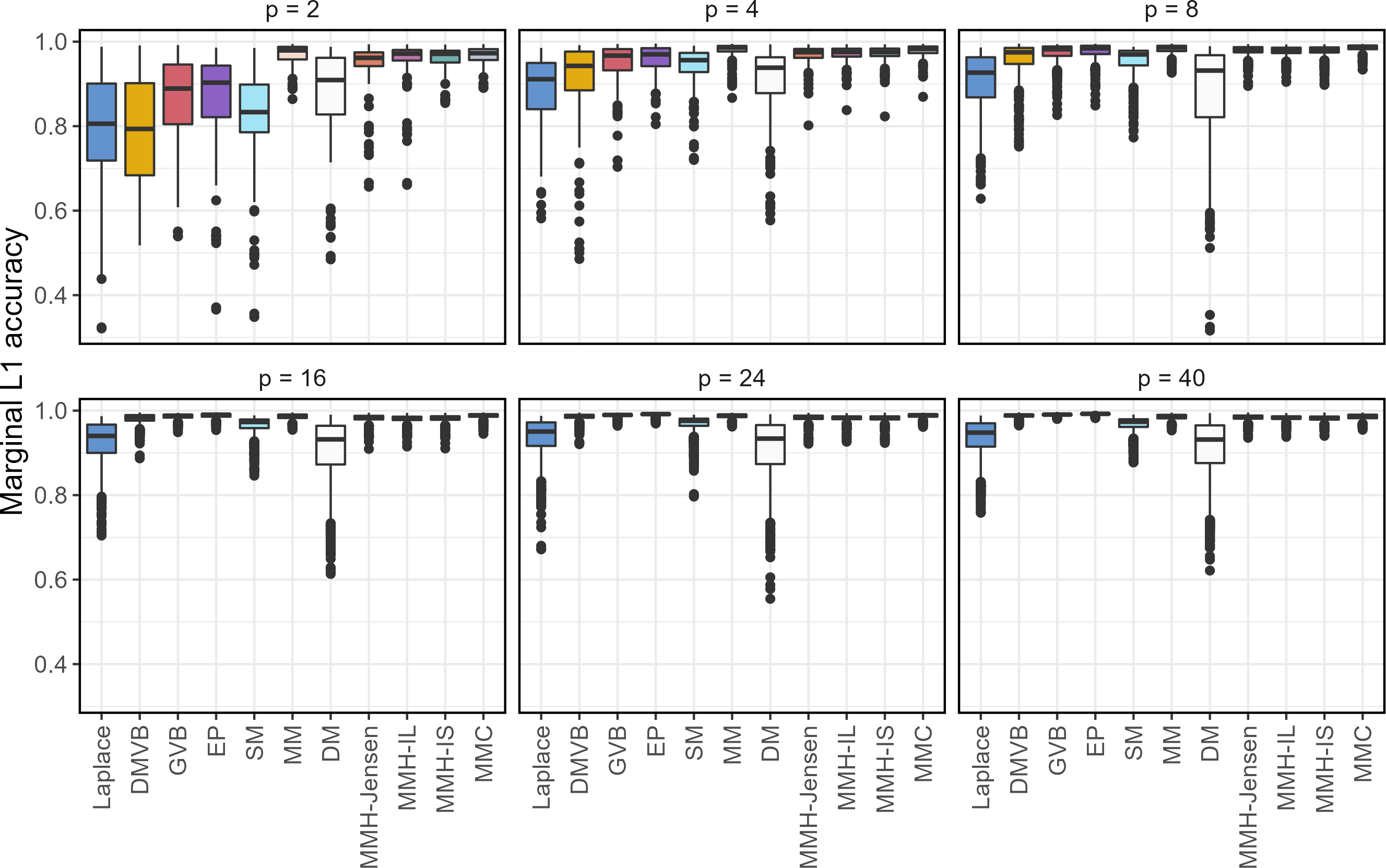}
    \end{center}
    \caption{Performance of approximation methods on simulated AR1 data across different dimensions for probit regression, with $n=4p$ (box plot).}
    \label{STA-sim-PR-4B-box}
\end{figure}

\begin{figure}
    \begin{center}
        \includegraphics[scale=0.58]{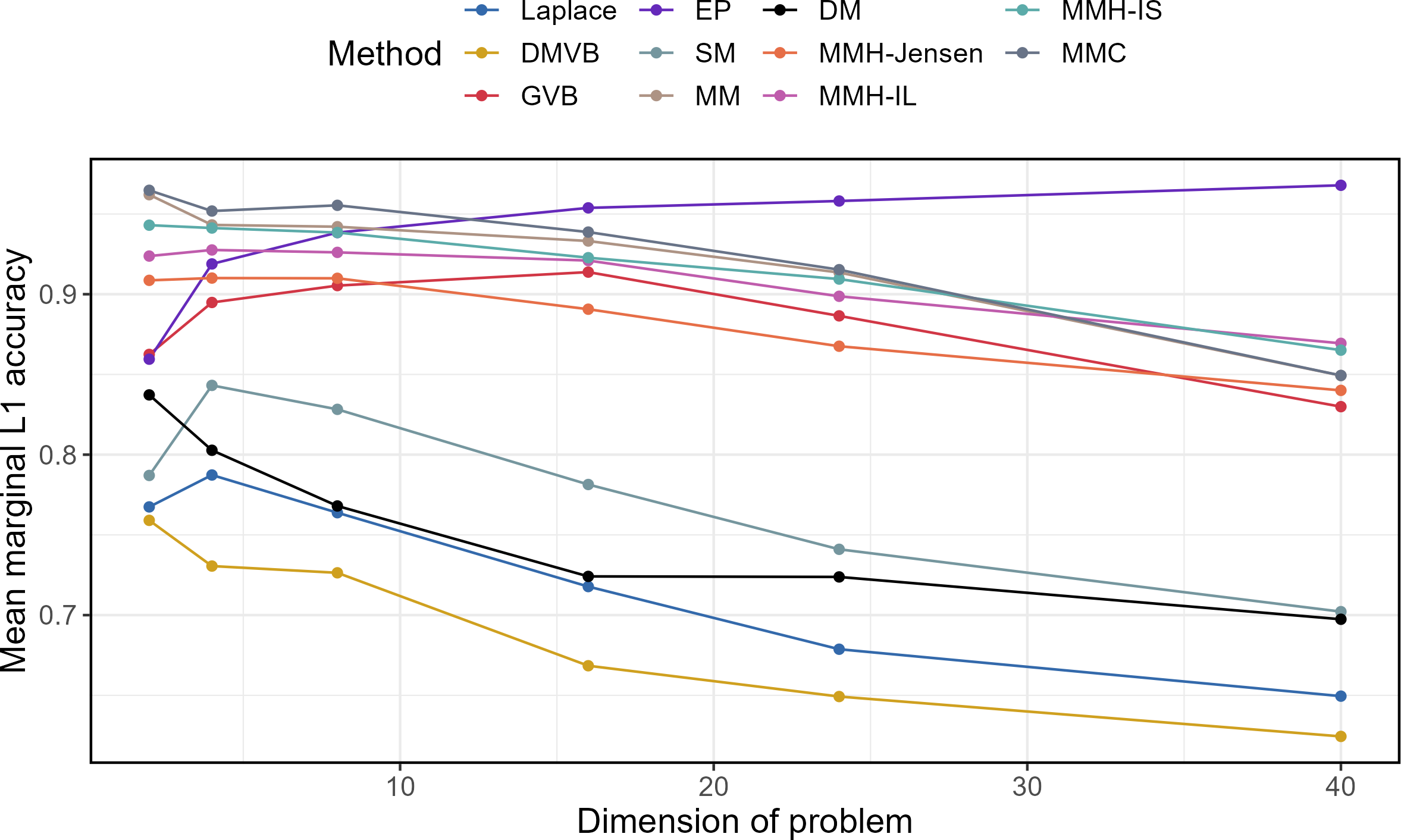}
    \end{center}
    \caption{Performance of approximation methods on simulated AR1 data across different dimensions for probit regression, with $n=2p$ (line graph).}
    \label{STA-sim-PR-2B-line}
\end{figure}

\begin{figure}
    \begin{center}
        \includegraphics[scale=0.58]{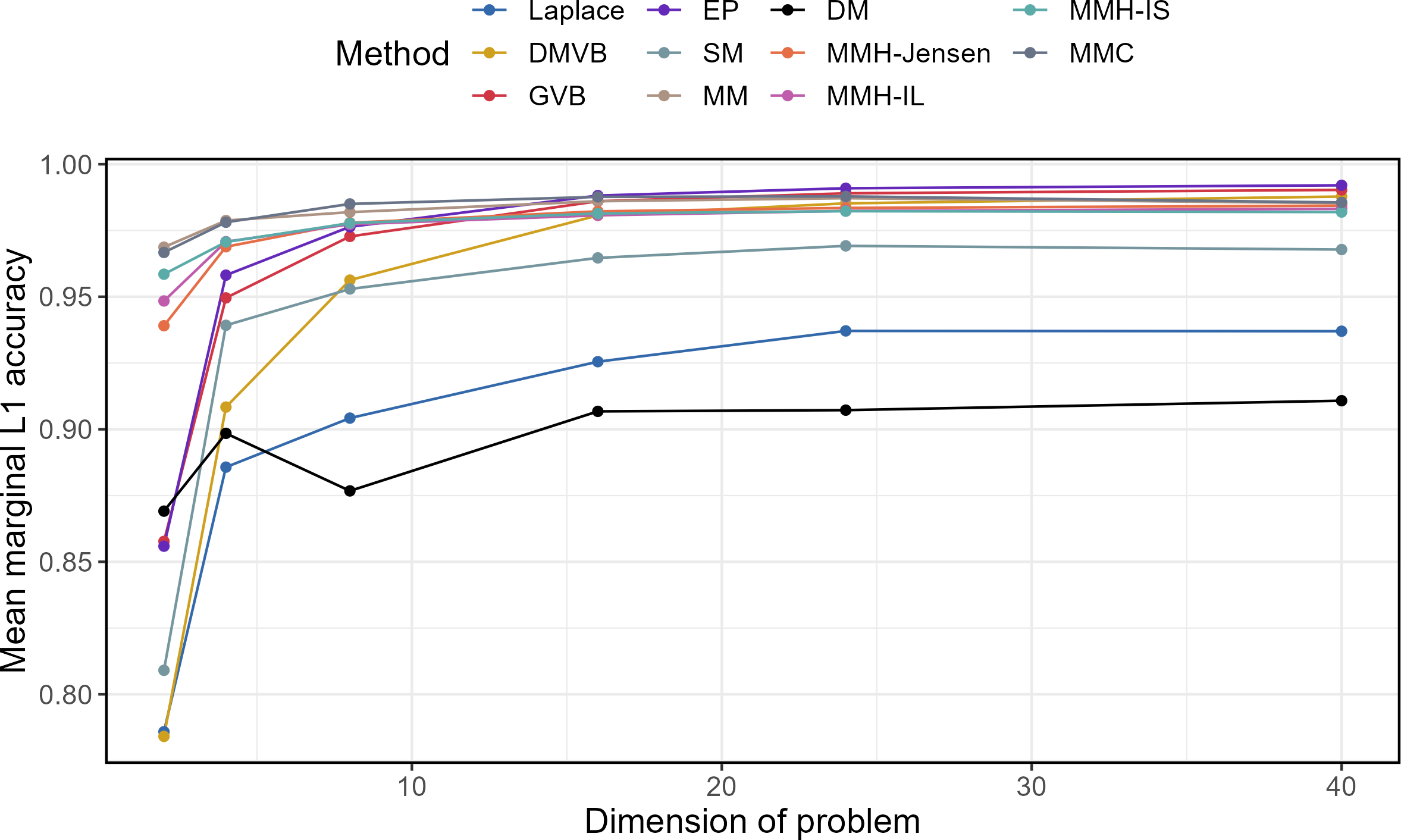}
    \end{center}
    \caption{Performance of approximation methods on simulated AR1 data across different dimensions for probit regression, with $n=4p$ (line graph).}
    \label{STA-sim-PR-4B-line}
\end{figure}

\begin{figure}
    \begin{center}
        \includegraphics[scale=0.62]{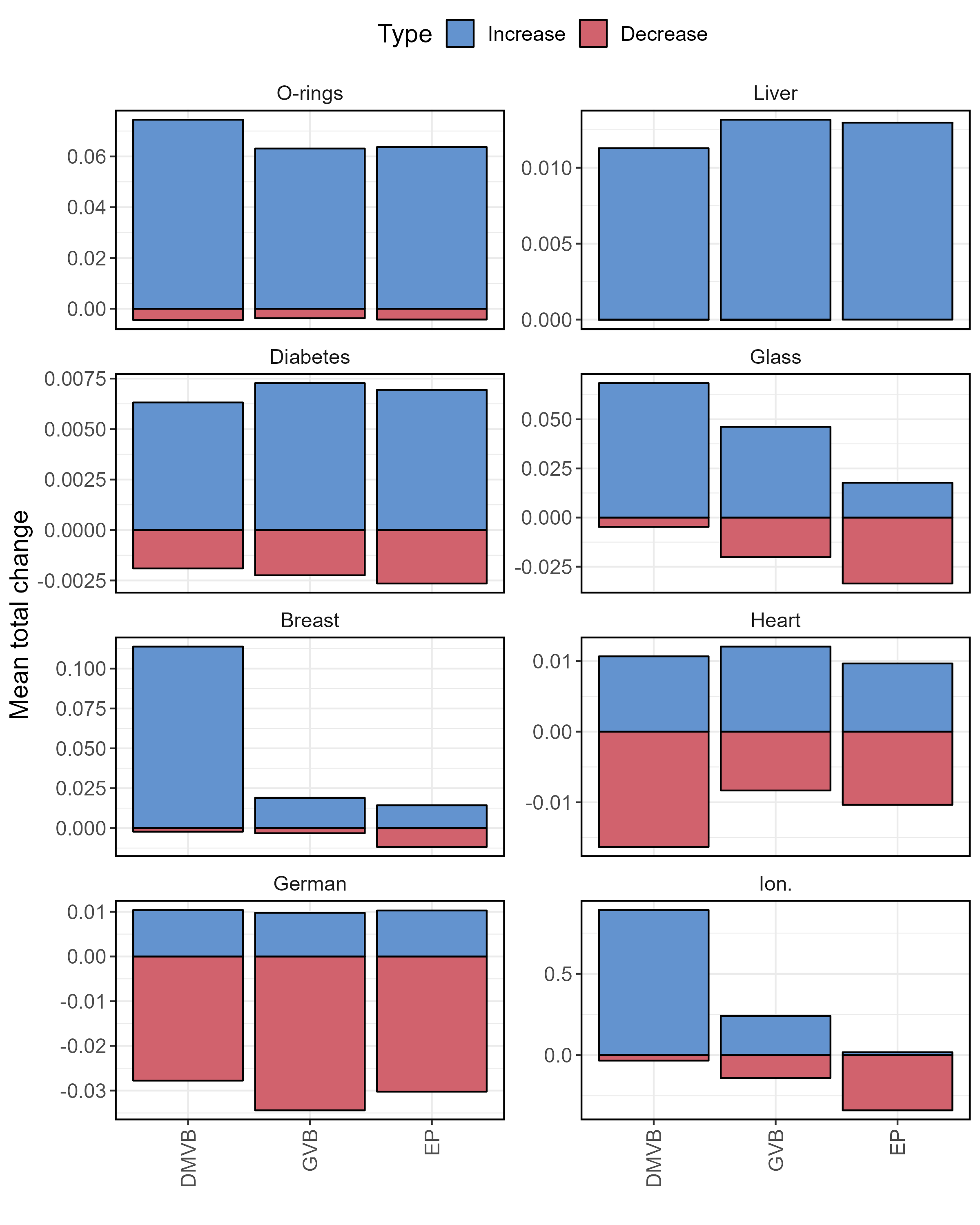}
    \end{center}
    \caption{Total change in marginal $L^1$ accuracy of a mean-mode-Hessian adjustment on benchmark datasets under probit regression.}
    \label{PHA-bench-PR-MMH-PH-dbar}
\end{figure}

\begin{figure}
    \begin{center}
        \includegraphics[scale=0.62]{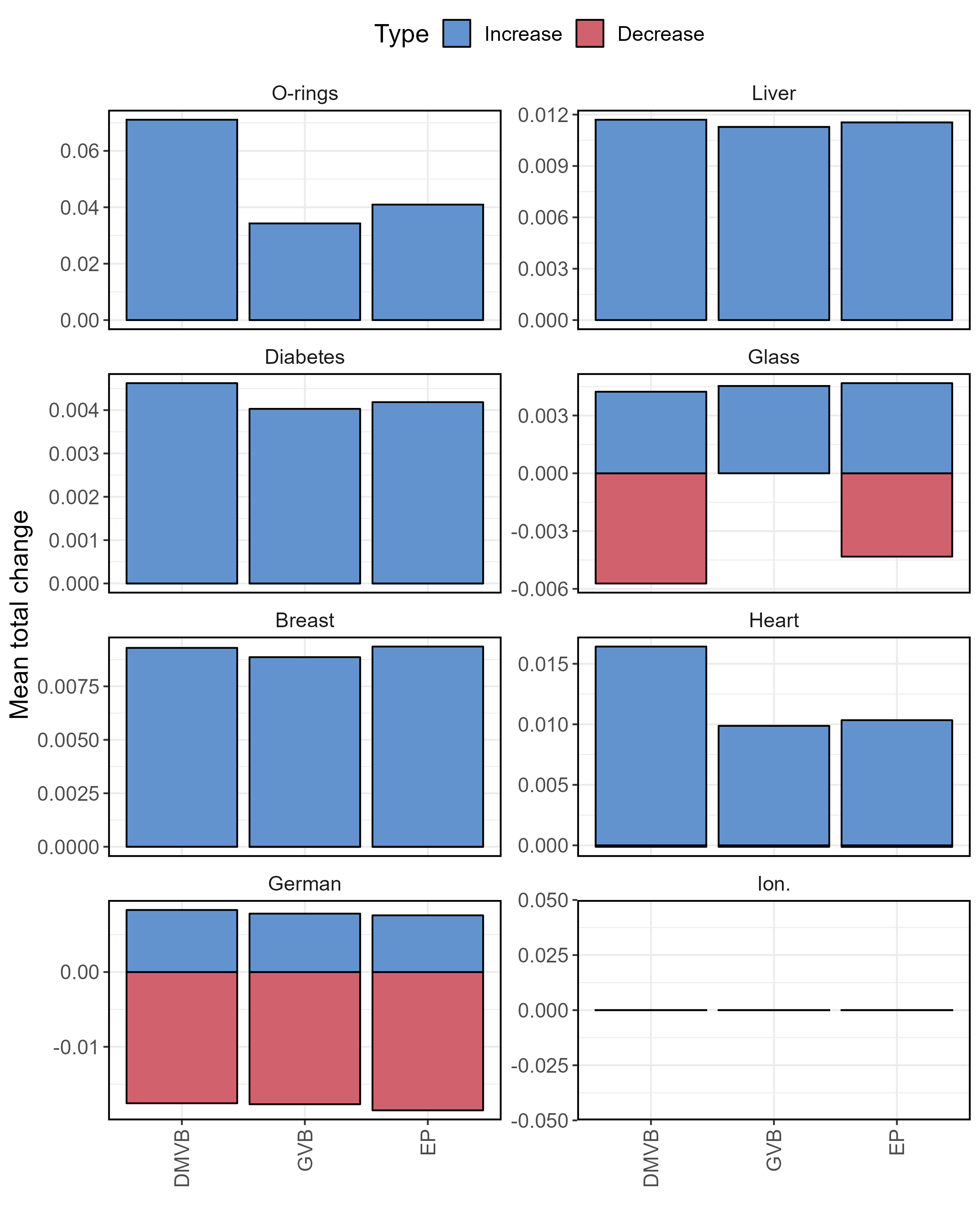}
    \end{center}
    \caption{Total change in marginal $L^1$ accuracy of a mean-mode-covariance adjustment on benchmark datasets under probit regression.}
    \label{PHA-bench-PR-MMC-PH-dbar}
\end{figure}

Additional plots corresponding to the probit regression results of the main text are provided.
These include box plots of marginal $L^1$ accuracies and line plots of mean marginal $L^1$ accuracies for each method, and are presented in Figures \ref{STA-sim-PR-2A-box} to \ref{STA-bench-PR-time}.
Similar plots corresponding to the dependent AR1 simulations are also presented in Figures \ref{STA-sim-PR-2B-box} to \ref{STA-sim-PR-4B-line}.
The relative performance of skew-normal matching here was similar to the independent covariate case.
Finally, post-hoc benchmark plots are given in Figures \ref{PHA-bench-PR-MMH-PH-dbar}
and \ref{PHA-bench-PR-MMC-PH-dbar}.

\section{Logistic regression supplement}
\label{sec:logistic-supplement}

\begin{figure}
    \begin{center}
        \includegraphics[scale=0.58]{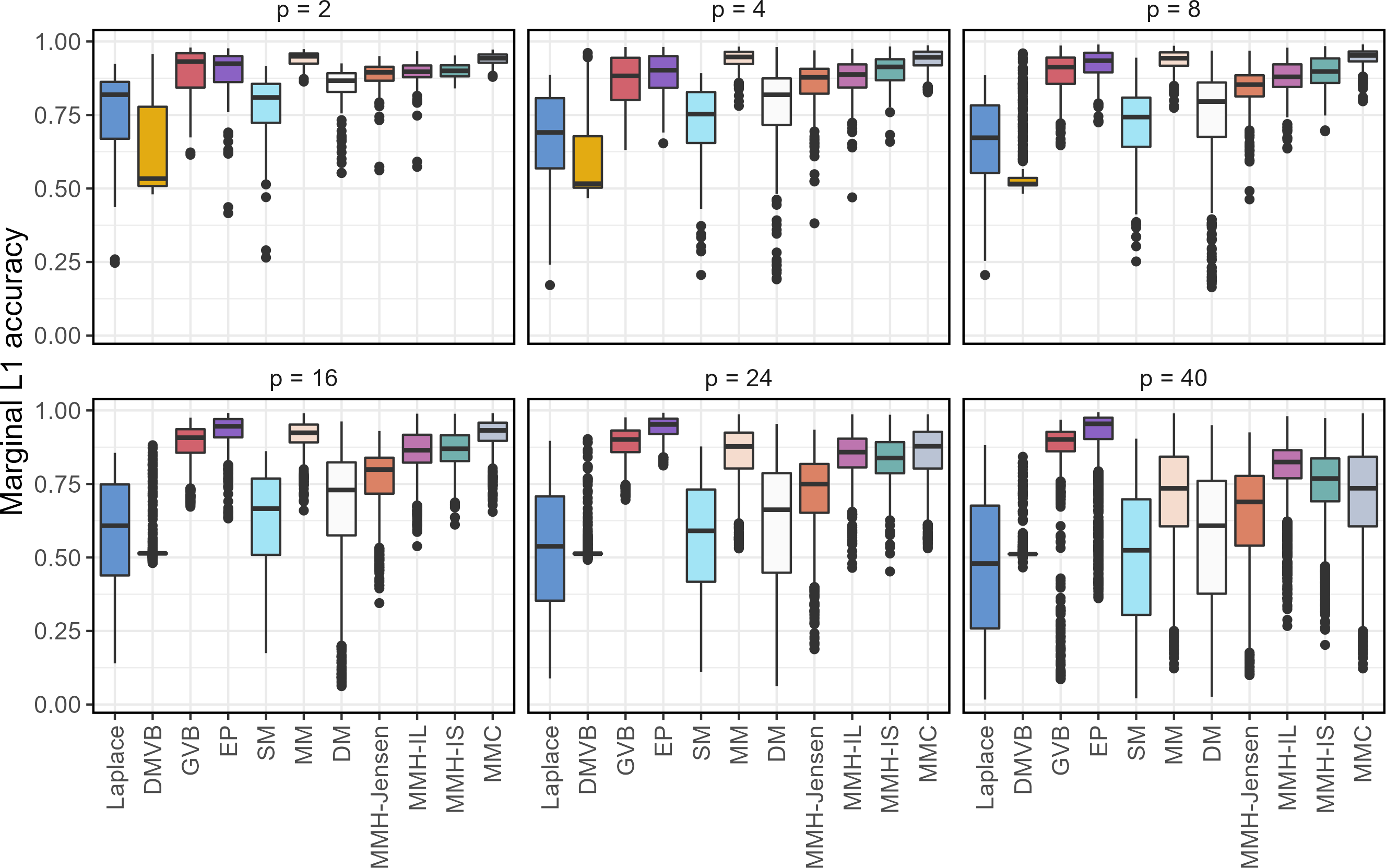}
    \end{center}
    \caption{Performance of approximation methods on simulated independent data across different dimensions for logistic regression, with $n=2p$ (box plot).}
    \label{STA-sim-LR-2A-box}
\end{figure}

\begin{figure}
    \begin{center}
        \includegraphics[scale=0.58]{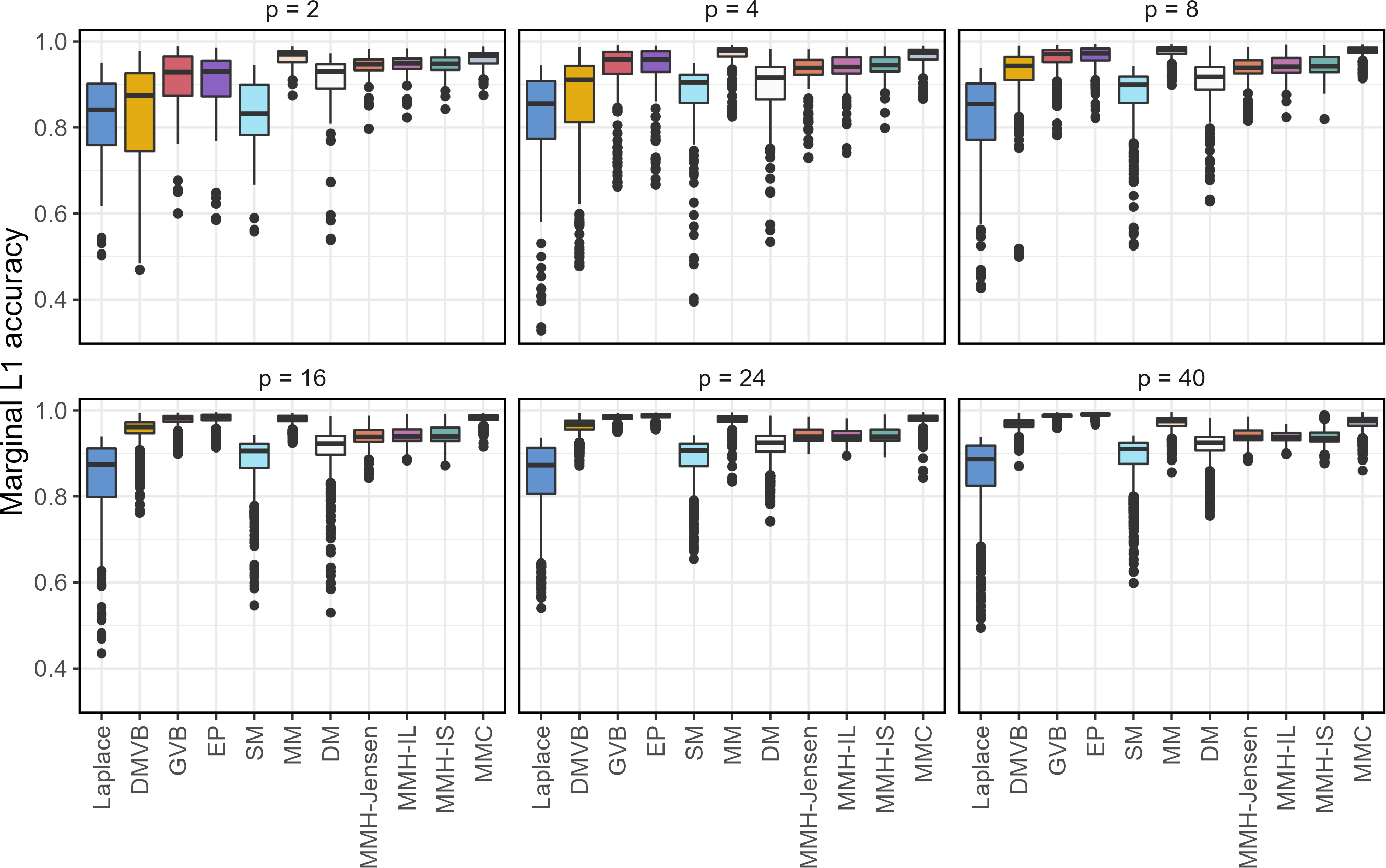}
    \end{center}
    \caption{Performance of approximation methods on simulated independent data across different dimensions for logistic regression, with $n=4p$ (box plot).}
    \label{STA-sim-LR-4A-box}
\end{figure}

\begin{figure}
    \begin{center}
        \includegraphics[scale=0.58]{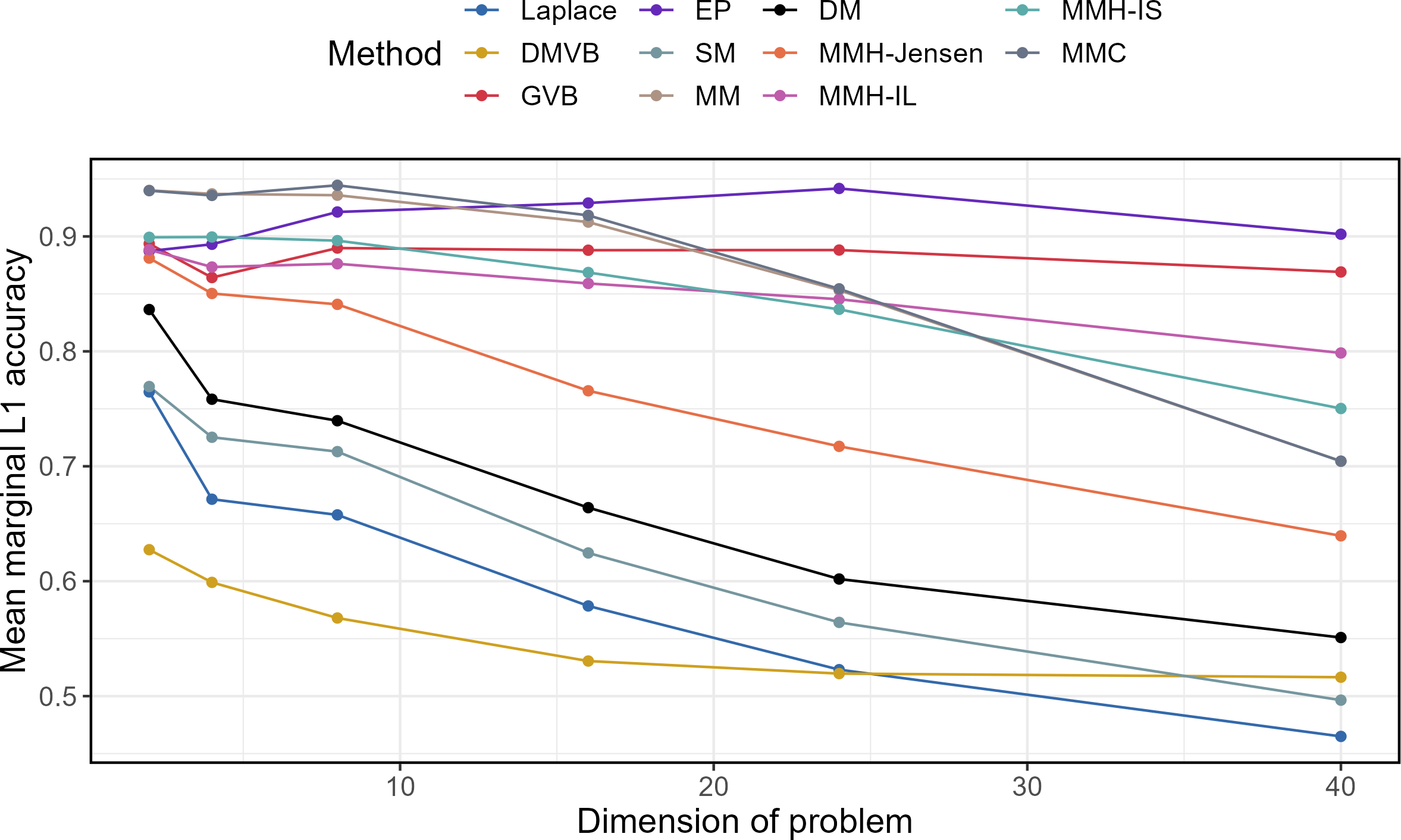}
    \end{center}
    \caption{Performance of approximation methods on simulated independent data across different dimensions for logistic regression, with $n=2p$ (line graph).}
    \label{STA-sim-LR-2A-line}
\end{figure}

\begin{figure}
    \begin{center}
        \includegraphics[scale=0.58]{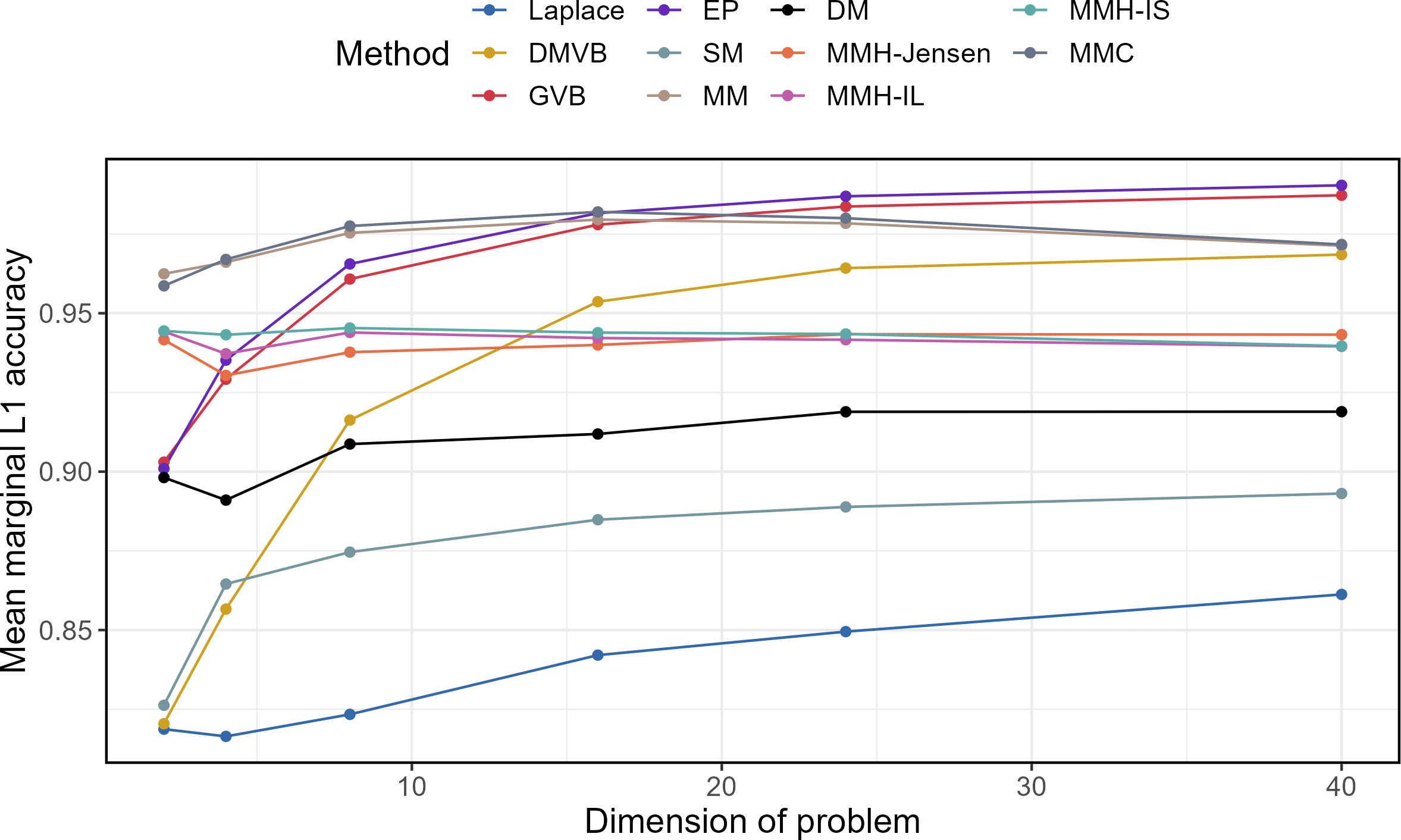}
    \end{center}
    \caption{Performance of approximation methods on simulated independent data across different dimensions for logistic regression, with $n=4p$ (line graph).}
    \label{STA-sim-LR-4A-line}
\end{figure}

\begin{figure}
    \begin{center}
        \includegraphics[scale=0.58]{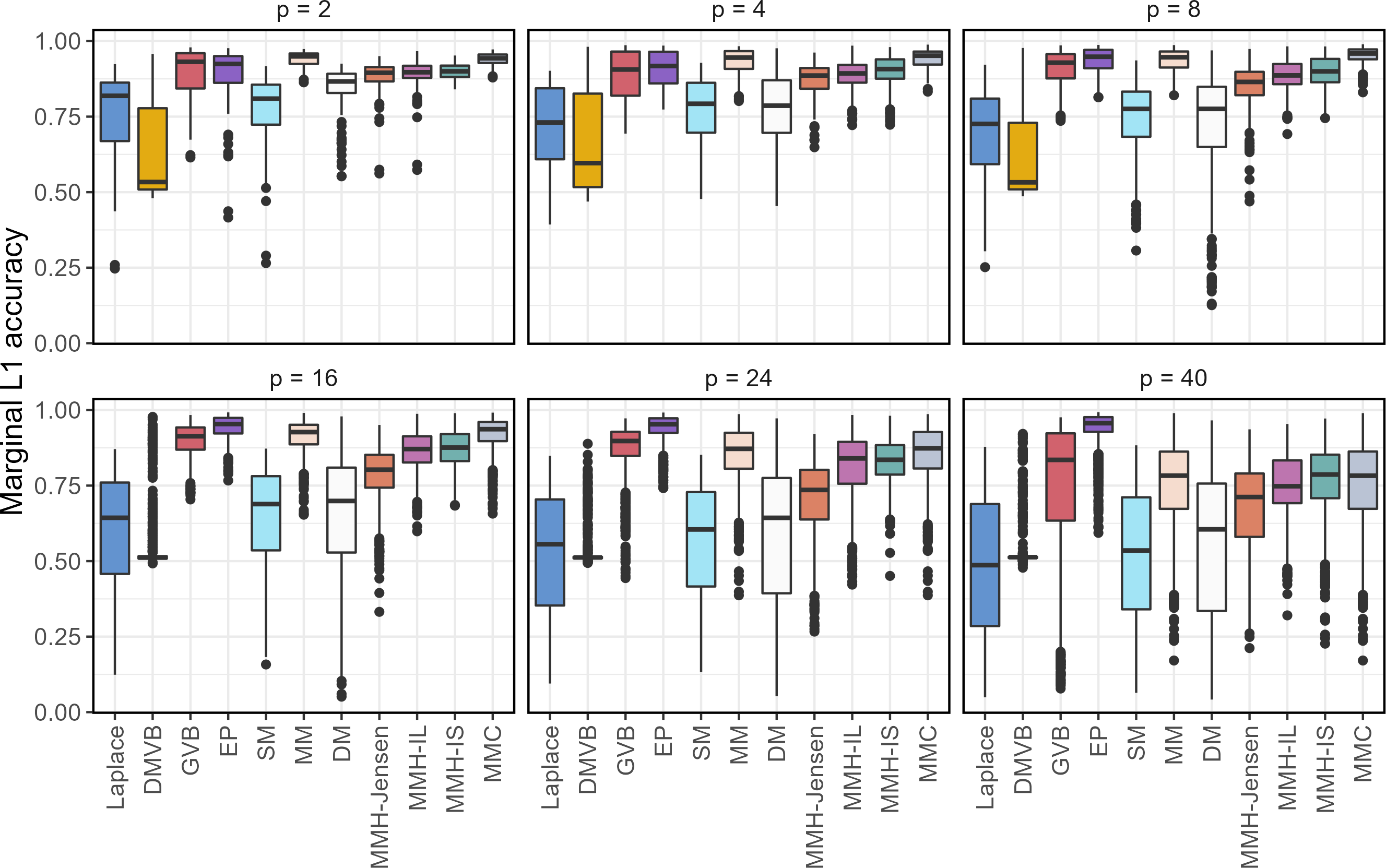}
    \end{center}
    \caption{Performance of approximation methods on simulated AR1 data across different dimensions for logistic regression, with $n=2p$ (box plot).}
    \label{STA-sim-LR-2B-box}
\end{figure}

\begin{figure}
    \begin{center}
        \includegraphics[scale=0.58]{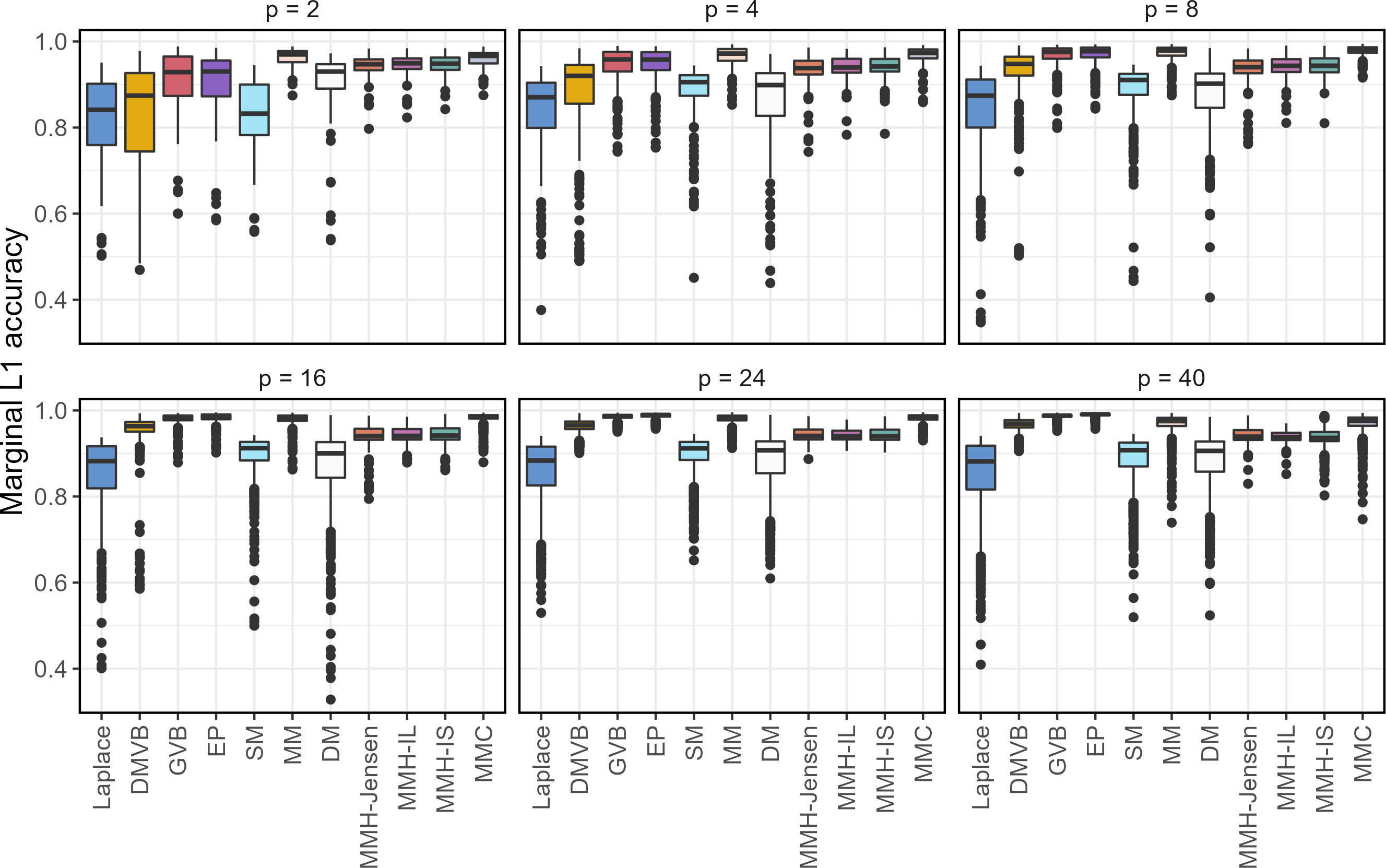}
    \end{center}
    \caption{Performance of approximation methods on simulated AR1 data across different dimensions for logistic regression, with $n=4p$ (box plot).}
    \label{STA-sim-LR-4B-box}
\end{figure}

\begin{figure}
    \begin{center}
        \includegraphics[scale=0.58]{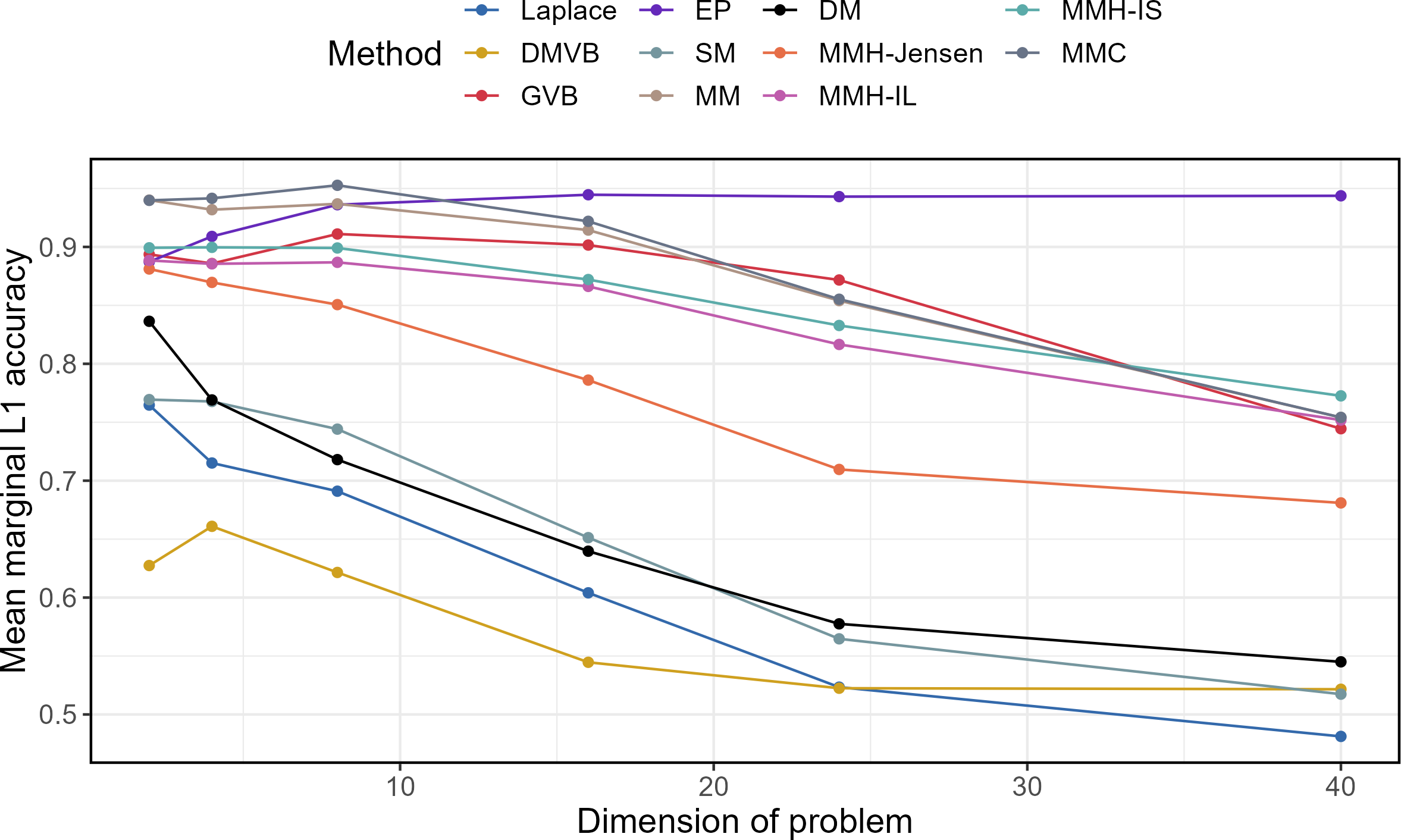}
    \end{center}
    \caption{Performance of approximation methods on simulated AR1 data across different dimensions for logistic regression, with $n=2p$ (line graph).}
    \label{STA-sim-LR-2B-line}
\end{figure}

\begin{figure}
    \begin{center}
        \includegraphics[scale=0.58]{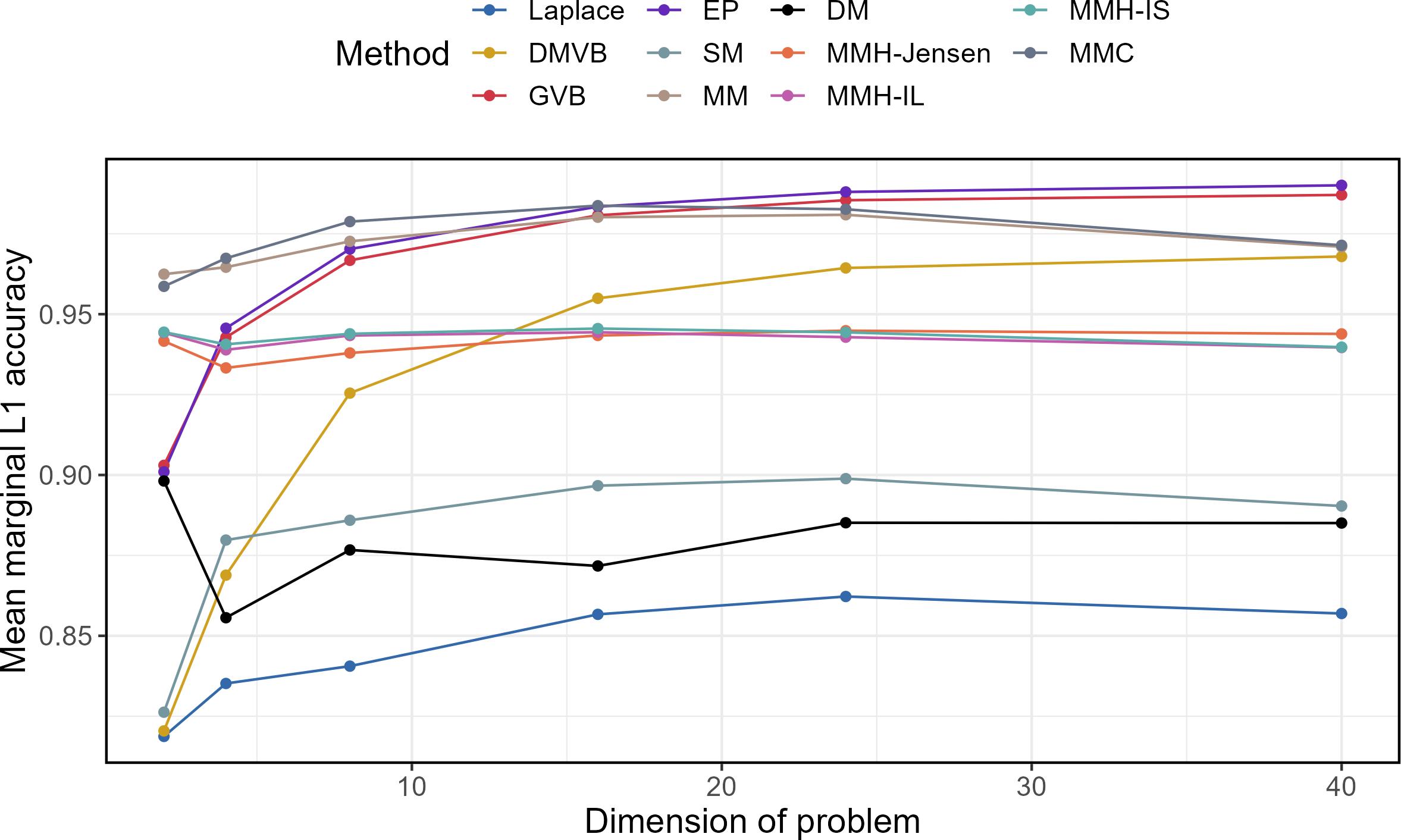}
    \end{center}
    \caption{Performance of approximation methods on simulated AR1 data across different dimensions for logistic regression, with $n=4p$ (line graph).}
    \label{STA-sim-LR-4B-line}
\end{figure}

\begin{figure}
    \begin{center}
        \includegraphics[scale=0.65]{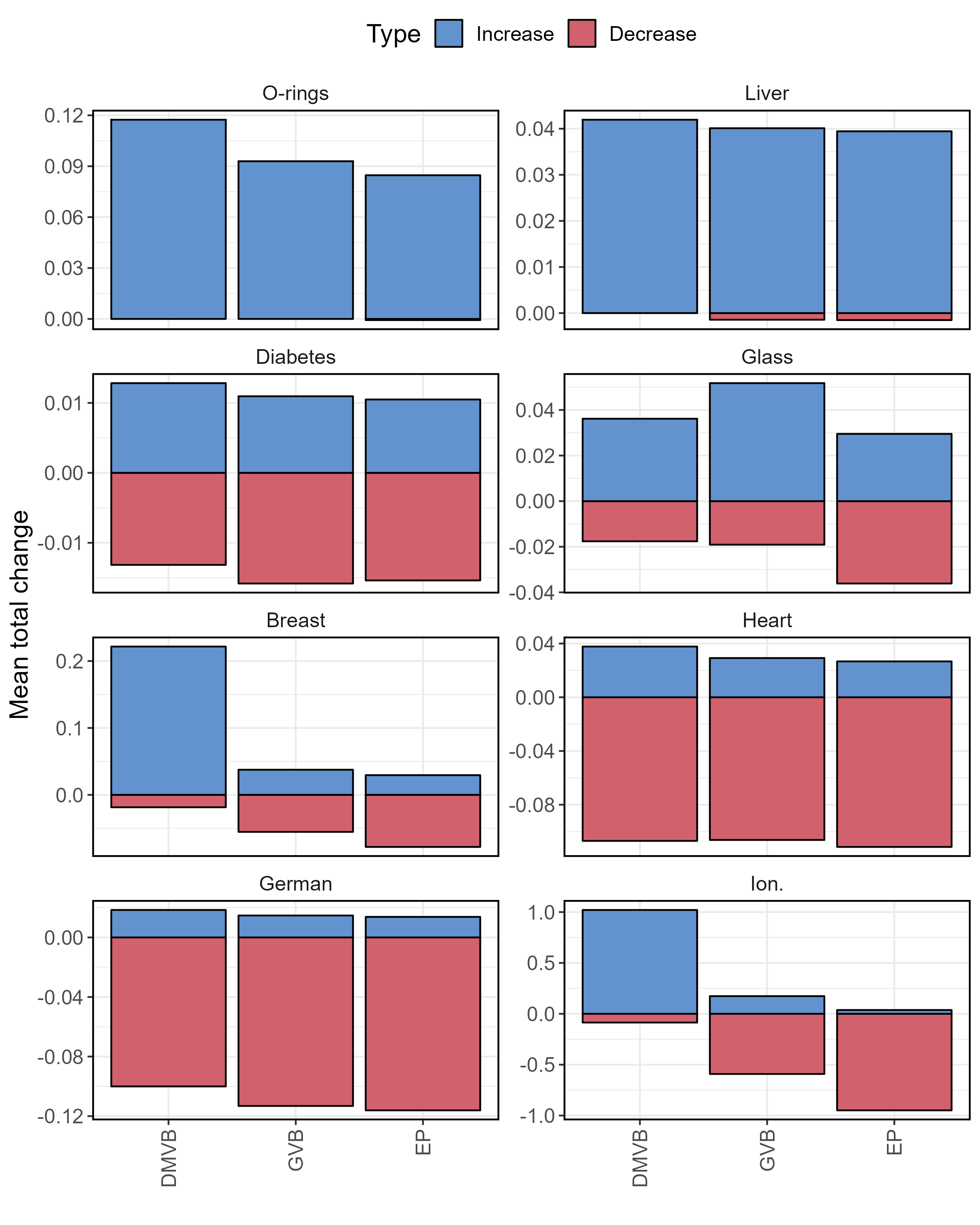}
    \end{center}
    \caption{Total change in marginal $L^1$ accuracy of a mean-mode-Hessian adjustment on benchmark datasets under logistic regression.}
    \label{PHA-bench-LR-MMH-PH-dbar}
\end{figure}

\begin{figure}
    \begin{center}
        \includegraphics[scale=0.65]{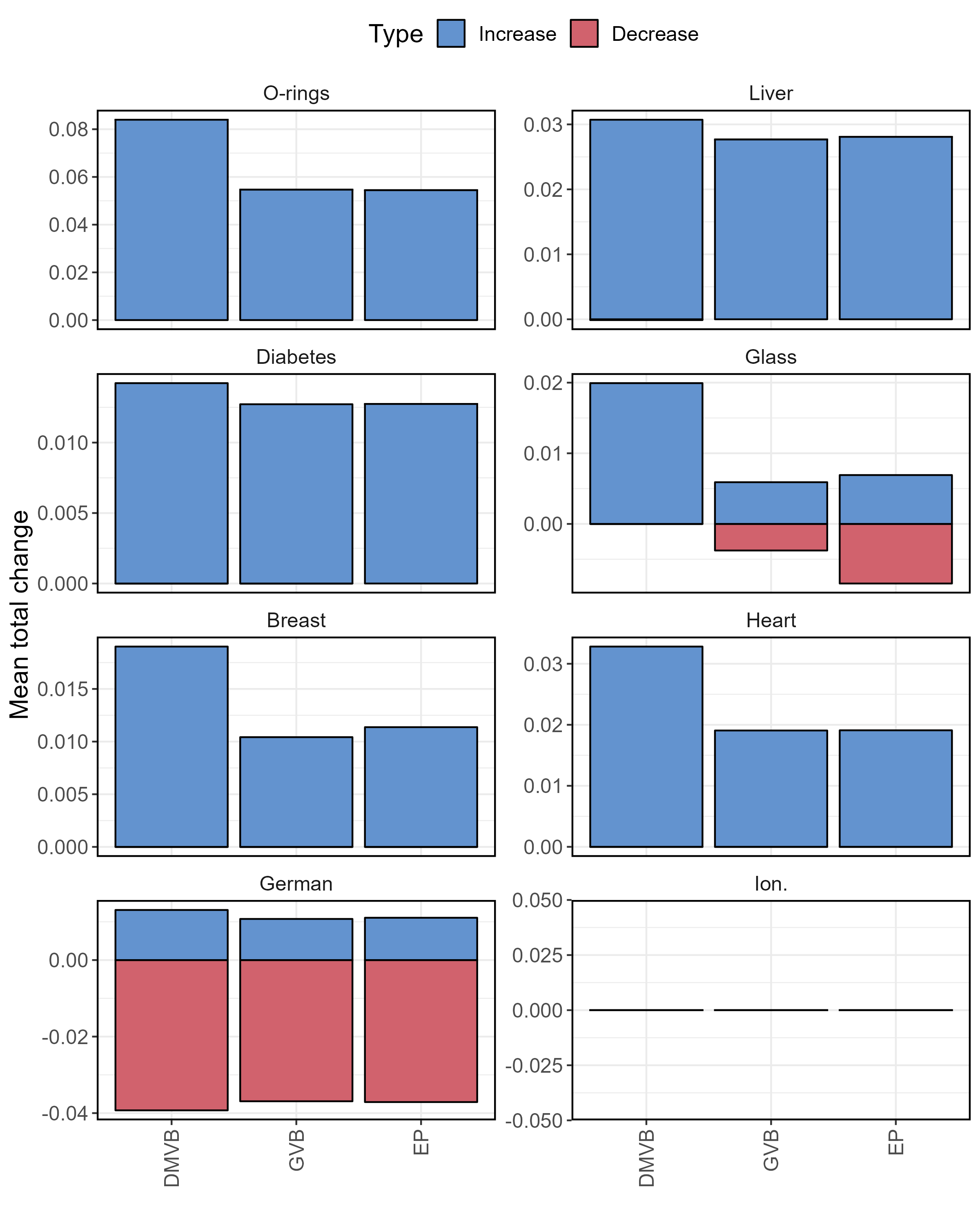}
    \end{center}
    \caption{Total change in marginal $L^1$ accuracy of a mean-mode-covariance adjustment on benchmark datasets under logistic regression.}
    \label{PHA-bench-LR-MMC-PH-dbar}
\end{figure}

Plots corresponding to the logistic regression case, where the skew-normal matching method was used as a standalone approximation, are provided.
These include box plots of marginal $L^1$ accuracies, line plots of mean marginal $L^1$ accuracies, and line plots of average time taken for each method, and are presented in Figures \ref{STA-sim-LR-2A-box} to \ref{STA-sim-LR-4A-line}.
The results are comparable to the probit regression case.
Similar plots corresponding to the dependent AR1 simulations are also presented in Figures \ref{STA-sim-LR-2B-box} to \ref{STA-sim-LR-4B-line}.
The relative performance of skew-normal matching here was similar to the independent covariate case.
Finally, post-hoc benchmark plots are given in Figures \ref{PHA-bench-LR-MMH-PH-dbar}
and \ref{PHA-bench-LR-MMC-PH-dbar}.

\bibliographystyle{plainnat}
\bibliography{bib/main}